\documentclass[10pt,journal]{IEEEtran}

\usepackage{amssymb}
\usepackage{amsmath}
\usepackage{cite}
\usepackage{url}
\usepackage{xcolor}
\usepackage{cite,graphicx,amsmath,amssymb}
\usepackage{fancyhdr}
\usepackage{mdwmath}
\usepackage{mdwtab}
\usepackage{caption}
\usepackage{amsthm}
\usepackage{setspace}
\usepackage{hyperref}
\usepackage{algorithm}
\usepackage{algorithmic}
\usepackage{pdfpages}
\usepackage{mathtools}
\usepackage{subcaption}
\usepackage{bbm}
\usepackage{stfloats}
\usepackage{balance}

\hypersetup{colorlinks=true,
linkcolor=blue,
citecolor=blue,      
urlcolor=black,
}

\graphicspath{{./src/img/}}

\DeclareMathOperator*{\argmax}{argmax}
\DeclareMathOperator*{\argmin}{argmin}

\newtheorem{remark}{Remark}
\newtheorem{theorem}{Theorem}

\newtheorem{lemma}{Lemma}

\newtheorem{corollary}{Corollary}

\allowdisplaybreaks
\title{Performance Analysis of Near-Field Sensing in Wideband MIMO Systems}

\author{
        Zhaolin~Wang,~\IEEEmembership{Member,~IEEE,}
        Xidong~Mu,~\IEEEmembership{Member,~IEEE,} \\
        and Yuanwei~Liu,~\IEEEmembership{Fellow,~IEEE}
\thanks{A early version of this paper will be presented in part at the IEEE Global Communications Conference, Cape Town, South Africa, 2024 \cite{conference_version}. Zhaolin Wang is with the School of Electronic Engineering and Computer Science, Queen Mary University of London, London E1 4NS, U.K. (e-mail: zhaolin.wang@qmul.ac.uk). Xidong Mu is with the Centre for Wireless Innovation (CWI), Queen's University Belfast, Belfast, BT3 9DT, U.K. (e-mail: x.mu@qub.ac.uk). Yuanwei Liu is with the Department of Electrical and Electronic Engineering, The University of Hong Kong, Hong Kong (e-mail: yuanwei@hku.hk).}
\vspace{-0.5cm}
}

\usepackage{pdfpages}

\begin{document}

\maketitle
\begin{abstract}
    \textcolor{black}{The performance of near-field sensing (NISE) in a legacy wideband multiple-input multiple-output (MIMO) orthogonal frequency-division multiplexing (OFDM) communication systems is analyzed. The maximum likelihood estimates (MLE) for the target’s distance and angle relative to the antenna array are derived.} To evaluate the estimation error, closed-form analytical expressions of Cramér-Rao bounds (CRBs) are derived for both uniform linear arrays (ULAs) and uniform circular arrays (UCAs). The asymptotic CRBs are then analyzed to reveal the scaling laws of CRBs with respect to key system parameters, including array size, bandwidth, and target distance. \textcolor{black}{Our results reveal that 1) the mean-squared error achieved by MLEs approaches CRBs in the high signal-to-noise ratio regime; 2) a larger array aperture does not necessarily improve NISE performance, especially with ultra-large bandwidth; 3) large bandwidth sets an estimation error ceiling for NISE as target distance increases; 4) array aperture and bandwidth, rather than the number of antennas and subcarriers, are the key factors affecting wideband NISE performance; and 5) UCAs offer superior, angle-independent wideband NISE performance compared to ULAs with the same aperture.}
\end{abstract}

\section{Introduction}

\IEEEPARstart{M}ULTIPLE-input multiple-output (MIMO) techniques have been pivotal in wireless communications and sensing systems \cite{bjornson2023twenty, fishler2004mimo}. By utilizing multiple antennas at transceivers, MIMO increases spatial degrees of freedom (DoFs), enhancing both communication rates and sensing accuracy. The similarity in hardware and signal processing between MIMO communication and sensing makes MIMO a key technology for integrated sensing and communication (ISAC), a crucial feature in next-generation wireless networks \cite{liu2022integrated, 3GPP_ISAC}. The current fifth-generation (5G) network, based on massive MIMO \cite{5595728}, incorporates numerous antennas to boost DoFs, energy efficiency, and resilience to hardware imperfections \cite{bjornson2014massive}. In 5G, a typical massive MIMO base station (BS) features 64 antennas, with an array aperture sufficiently compact to disregard the near-field effect. Therefore, the prior studies of massive MIMO focused mainly on the far-field scenarios \cite{el2014spatially, yu2016alternating, sohrabi2016hybrid}. 


In sixth-generation (6G) networks, the focus is shifting to extremely large-scale MIMO (XL-MIMO) systems \cite{zhang20196g, wang2023extremely}, which could feature hundreds or thousands of antennas. This significantly increases the array size, leading to fundamental changes in the electromagnetic properties of signals \cite{liu2023near_tutorial}. In particular, as the array size grows, the near-field region around BS can be greatly expanded, especially in higher frequency bands like millimeter-wave and terahertz. Unlike far-field scenarios, where signals propagate as planar waves, near-field signals exhibit spherical wave propagation, necessitating a new perspective on communication and sensing performance.


In this paper, we focus on the performance of Near-Field Sensing (NISE). From the sensing perspective, traditional far-field sensing can only resolve the angular information of signals in the spatial domain, while the estimation of a target's distance from the BS generally relies on the system bandwidth \cite{liu2022integrated, wang2023rethinking}. This implies that a substantial bandwidth is crucial for accurately estimating distance. Conversely, NISE benefits from spherical wave propagation, enabling the antenna array to discern both the wave's direction and its travel distance \cite{liu2023near_tutorial}. This capability facilitates the spatial-domain estimation of the target's distance, thereby diminishing the dependency on extensive bandwidth. Therefore, most of the existing research on NISE focus on its fundamental limits in narrowband systems
\cite{el2010conditional, 9439203, 6362262, wang2023cram, 7981398, 9508850, 9950340, 10147356}. The Cramér-Rao bound (CRB) is a widely used mathematical method to evaluate sensing performance since it provides a tractable and tight lower bound of unbiased estimators under some mild and general conditions. In particular, the conditional and unconditional CRBs of narrowband NISE for mono-static and bi-static setups were derived in \cite{el2010conditional} and \cite{9439203}, respectively.
As a further advance, the authors of \cite{6362262} conducted a comparison of various deterministic performance bounds in mono-static narrowband NISE systems. However, the CRBs derived in the above work rely either on the Fresnel approximation for the near-field channel or on the numerical calculation of partial derivatives, making it difficult to avoid approximation errors or to gain insights into system design. Thus, the authors of \cite{wang2023cram} first derived the closed-form CRBs for narrowband NISE based on exact near-field channel models and carried out the asymptotic analysis with respect to some key system parameters. Furthermore, the CRBs of NISE in wideband multi-carrier systems were derived in \cite{7981398}, whose accuracy was validated through Monte-Carlo simulations. The authors of \cite{9508850} studied the CRBs of NISE in a scenario with moving targets, considering both uniform linear arrays (ULAs) and uniform circular arrays (UCAs), but focusing on the simple narrowband scenarios. Recently, there have also been some research works to study the CRBs of narrowband NISE based on electromagnetic theory, such as \cite{9950340} and \cite{10147356}. However, the fundamental sensing performance characterization of NISE in wideband systems has not been explored.

\subsection{Motivation and Contributions}
As discussed above, the key difference between NISE and far-field sensing lies in their approaches to distance estimation. NISE can directly measure distance within the spatial domain, in contrast to far-field sensing, which depends on extensive bandwidth for accurate distance estimation. This leads to an intriguing question: \emph{How does the large bandwidth impact the performance of NISE?} While this question was initially explored in \cite{7981398}, the derived CRBs therein lack closed-form expressions, which makes it hard to directly understand and quantify the relationship between sensing performance and crucial system parameters. Additionally, most existing works on NISE focus primarily on ULAs or uniform planar arrays. However, recent studies suggested that UCAs have the potential to significantly expand the near-field region, offering promising improvements to NISE performance \cite{wu2023enabling, 9508850, wang2023rethinking}. Driven by the above considerations and the growing trend towards the ISAC paradigm, this paper studies the performance of NISE within a conventional wideband communication system, exploring the implications for both ULAs and UCAs. The main contributions of this paper are summarized as follows:
\begin{itemize}
    \item We investigate the performance of NISE in a legacy wideband MIMO communication system employing orthogonal frequency-division multiplexing (OFDM). We present signal and channel models for NISE that account for both near-field and wideband effects. Based on these models, we derive maximum likelihood estimators for the direction and distance of the target.
    \item We derive closed-form analytical expressions of CRBs for NISE in wideband MIMO-OFDM systems with ULAs and UCAs, respectively. Based on these expressions, we analyze the asymptotic CRB performance to unveil the impact of key system parameters, such as array size, bandwidth, and field region, on NISE performance.
    \item We validate the analytical results and the effectiveness of the maximum likelihood estimates through numerical results. Both analytical and numerical results show that: 1) A larger array aperture does not always lead to better NISE performance, especially when the bandwidth is ultra-large; 2) Large bandwidth provides an estimation error ceiling for NISE as the target's distance increases; 3) It is the array aperture and bandwidth, rather than the number of antennas and subcarriers, that fundamentally change the performance of wideband NISE; 4) The UCA provides better, angle-independent wideband NISE performance compared to the ULA with the same aperture.
\end{itemize}

\subsection{Organization and Notations}

The remainder of this paper is structured as follows: Section \ref{sec:system_model} introduces the system model for wideband NISE. Section \ref{sec:MLE} derives the maximum likelihood estimation method for wideband NISE. Section \ref{sec:analysis} delves into the performance analysis of wideband NISE. Section \ref{sec:results} discusses the numerical results. Finally, Section \ref{sec:conclusion} concludes the paper.

\emph{Notations:} Scalars, vectors, and matrices are represented
by lower-case, bold-face lower-case, and bold-face upper-case letters, respectively. The transpose, conjugate, conjugate transpose, trace, determinant, and Frobenius norm of matrix $\mathbf{X}$ are denoted by $\mathbf{X}^T$, $\mathbf{X}^*$, $\mathbf{X}^H$, $\mathrm{tr}(\mathbf{X})$, $\mathrm{det}(\mathbf{X})$, and $\|\mathbf{X}\|_F$, respectively. $\mathcal{CN}(0, \sigma^2)$ circularly symmetric complex Gaussian random distribution with zero mean and variance $\sigma^2$. $\mathbf{I}_{N}$ denotes an $N \times N$ identity matrix. $\Re(\cdot)$ and $\Im(\cdot)$ denote the real and imaginary parts of a complex number. $\mathbb{E}[\cdot]$ denotes the e statistic expectation of a random variable. 

\section{System Model} \label{sec:system_model}
\begin{figure}[t!]
    \centering
    \includegraphics[width=0.3\textwidth]{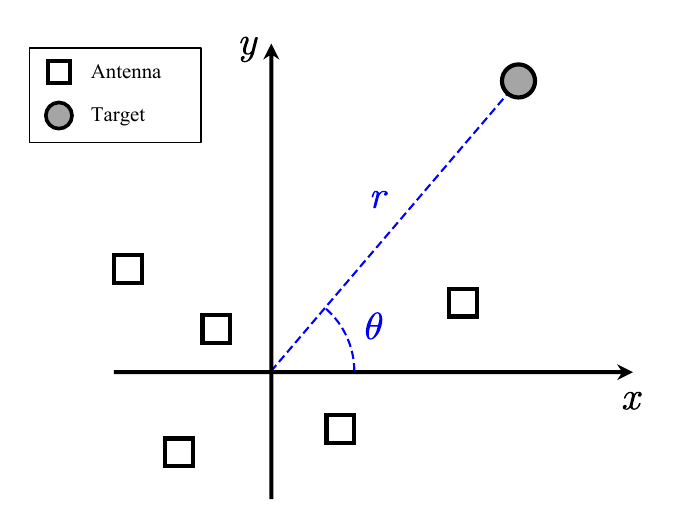}
    \caption{Geometry of the considered system.}
    \label{fig_system_model}
\end{figure}

\textcolor{black}{In this study, we investigate the performance of NISE in a legacy wideband MIMO-OFDM communication system employing a dual-functional $N$-antenna BS.}
At the BS, we assume a shared antenna array for transmitting and receiving through the use of circulators and the perfect self-interference cancellation through the full-duplex techniques \cite{6832464}. Furthermore, the communication users carry out the standard OFDM receive operations. This work will focus on the sensing aspect. As depicted in Fig. \ref{fig_system_model}, we examine the simplest scenarios including a point-like target in a two-dimensional coordinate system under the near-field channel model. \textcolor{black}{The target is assumed to either have a fixed position or a very low velocity, resulting in negligible Doppler frequencies.}\footnote{\textcolor{black}{In contrast to far-field systems, the Doppler frequency in near-field systems exhibits significantly different characteristics, as it is influenced by both the radial and transverse velocity of the targets \cite{10664591}. Analyzing the impact of near-field Doppler frequency on sensing performance is beyond the scope of this paper and will be addressed in future work.}} The angle and distance of the target with respect to the origin of the coordinate system are denoted by $\theta$ and $r$, respectively.

\subsection{Transmit Signal Model}

Consider an OFDM frame with $L$ OFDM symbols. Let $f_c$ denote the carrier frequency, $M$ denote the number of subcarriers, $T_{\mathrm{s}}$ denote the elementary duration of an OFDM symbol, and $T_{\mathrm{cp}}$ denote the duration of the CP. Consequently, the subcarrier spacing, the overall bandwidth, and the overall symbol duration of the OFDM system are $\Delta f = \frac{1}{T_{\mathrm{s}}}$, $B = M \Delta f$, $T_{\mathrm{tot}} = T_{\mathrm{s}} + T_{\mathrm{cp}}$, respectively. Then, the baseband transmit signal over an OFDM frame can be expressed as \cite{sturm2011waveform}
\begin{equation} \label{chapter_2_transmit_signal}
    \bar{\mathbf{x}}(t) = \frac{1}{\sqrt{M}} \sum_{l = 0}^{L-1} \sum_{m=0}^{M-1} \bar{\mathbf{x}}_m(l) e^{j 2\pi \delta_{m} \Delta f (t - l T_{\mathrm{tot}})} \Pi \left( \frac{t - l T_{\mathrm{tot}}}{T_{\mathrm{tot}}} \right),
\end{equation} 
where $\bar{\mathbf{x}}_m(l) \in \mathbb{C}^{N \times 1}$ denotes the signal on the $m$-th subcarriers in the $l$-th OFDM sysmbol, $\Pi (t)$ denotes the rectangular function which has a value of $1$ if $t \in [0,1]$ and $0$ otherwise, and $\delta_m = \frac{2m-M+1}{2}$. \textcolor{black}{Although all the communication data signals can be reused for sensing purposes, an additional dedicated sensing signal is needed to achieve a full transmission degree of freedom to form an optimal sensing beam, leading to the following transmit signal on the $m$-th subcarrier \cite{9124713}:
\begin{equation} \label{transmit_signal_m}
    \bar{\mathbf{x}}_m(l) = \sum_{k=1}^K \mathbf{w}_{m,k} c_{m,k}(l) + \mathbf{s}_m(l).
\end{equation} 
Here, $K$ denotes the number of communication users, $\mathbf{w}_{m,k} \in \mathbb{C}^{N \times 1}$ is the beamformer to convey the data symbol $c_{m,k}(l)$ to the $k$-user on the $m$-th subcarrier, and $\mathbf{s}_m(l) \in \mathbb{C}^{N \times 1}$ is the dedicated sensing signal. It is assumed that the data symbols $c_{m,k}(l)$ are independent and identically distributed with a unit power, which is also independent of the dedicated sensing signal $\mathbf{s}_m(l)$. Consequently, the covariance matrix of the transmit signal on the $m$-th subcarrier can be calculated as $\bar{\mathbf{R}}_m = \mathbb{E}\left[\bar{\mathbf{x}}_m(l) \bar{\mathbf{x}}_m^H(l)  \right] =\sum_{k=1}^K \mathbf{w}_{m,k} \mathbf{w}_{m,k}^H + \mathbf{R}_s$, where $\mathbf{R}_s = \mathbb{E}\left[\mathbf{s}_m(l) \mathbf{s}_m^H(l)  \right]$ represents the covariance matrix of the dedicated sensing signal. } 

\textcolor{black}{In this study, we focus on a sensing-prior design for the transmit signal \cite{9724205}. Specifically, the transmit signal on each subcarrier is designed to form an optimal sensing beam. We assume a practical scenario that the target is not known a priori. In this case, to guarantee optimal worst-case performance of sensing, the transmit signal needs to be spatially white \cite{stoica2007probing}, i.e., 
\begin{equation} \label{isotropic_beam}
    \bar{\mathbf{R}}_m \triangleq \mathbb{E}\left[ \bar{\mathbf{x}}_m(l) \bar{\mathbf{x}}^H_m(l)\right] = \frac{P_m}{N} \mathbf{I}_N, \forall m,
\end{equation}
where $P_m$ denotes power allocated to the $m$-th subcarrier. Under the above constraints on the covariance matrix, the transmit signal in \eqref{transmit_signal_m} can be further optimized to maximize the communication performance. We refer to \cite{9724205} for a comprehensive discussion about the corresponding system design and optimization method to enhance the communication performance on each subcarrier. In the sequel, we focus on the sensing performance under the condition \eqref{isotropic_beam}.}

\subsection{Receive Signal Model} \label{sec:receive_model}
Let $\mathbf{r} = [r \cos \theta, r \sin \theta]^T \in \mathbb{R}^2$ and $\mathbf{q}_n \in \mathbb{R}^2$ denote the coordinates of the target and the $n$-th antenna at the BS, respectively. The aperture of the antenna array at the BS is thus given by $D = \max_{i,j} \|\mathbf{q}_i - \mathbf{q}_j \|$.  According to the near-field model, the propagation distance from the $n$-th antenna to the target need to be calculated as the exact Euclidean distance $r_n = \| \mathbf{r} - \mathbf{q}_n \|$. We consider two typical antenna array geometries in sensing systems: ULAs and UCAs \cite{526899}. Let $d$ denote the antenna spacing. The origin of the coordinate system is put into the center of the antenna array. Then, the distance $r_n$ for these two array geometries as given as follows.
\begin{itemize}
    \item \emph{ULA:} We assume that the ULA is deployed along the $x$-axis. The coordinate of the $n$-th antenna is given by $\mathbf{q}_n = [\chi_n d, 0]^T$, with $\chi_n = n - \frac{N-1}{2}$. Therefore, the distance $r_n$ can be expressed as      
    \begin{equation} \label{ULA_distance}
        r_n = \sqrt{r^2 + \chi_n^2 d^2 - 2 r \chi_n d \cos \theta}.
    \end{equation}
 
    \item \emph{UCA:} For UCAs, the antennas are uniformly deployed on a circle with spacing $d$. The coordinate of the $n$-th antenna is given by $\mathbf{q}_n = [R \cos \psi_n, R \sin \psi_n]^T$, with $R = \frac{N d}{2 \pi}$ being the radius of the UCA and $\psi_n = \frac{2\pi n}{N}$. Therefore, the distance $r_n$ can be expressed as         
    \begin{equation} \label{distance_UCA}
        r_n = \sqrt{r^2 + R^2 - 2 r R \cos(\theta - \psi_n)}.
    \end{equation}
\end{itemize}

\textcolor{black}{Then, the round-trip propagation delay of the echo signal reflected by the target, from the $n$-th antenna to the $i$-th antenna at the BS, is given by 
\begin{equation}
    \tau_{n,i} = \frac{r_n + r_i}{c},
\end{equation}
where $c$ denotes the speed of light. Let $x_{m,n}(l) = [\mathbf{x}_m(l)]_n$ denote the baseband signals transmitted by the $n$-th antenna at the $m$-th subcarrier, the noiseless continuous-time baseband signal received at the $i$-th antenna at the BS can be modelled as \cite{goldsmith2005wireless, tse2005fundamentals}
\begin{align}
    &y_i(t) = \frac{1}{\sqrt{M}} \sum_{n=1}^{N} \sum_{l = 0}^{L-1} \sum_{m=0}^{M-1} \beta_{m, n, i} x_{m,n}(l) e^{j 2\pi \delta_{m} \Delta f \left(t - \tau_{n,i} - l T_{\mathrm{tot}}\right)} \nonumber \\
    &\hspace{2.7cm} \times e^{-j 2 \pi f_c \tau_{n,i}} \Pi \left( \frac{t - \tau_{n,i} - l T_{\mathrm{tot}}}{T_{\mathrm{tot}}} \right),
\end{align} 
where $\beta_{m,n,i}$ denote the frequency-dependent channel gain at the $m$-th subcarrier for the path from the $n$-th transmit antenna to the $i$-th receive antenna. According to the radar range equation \cite{richards2010principles}, $\beta_{m,n,i}$  can be modelled as 
\begin{equation}
    \beta_{m,n,i} = \frac{\sqrt{\epsilon_m} \beta_r}{r_n r_i}, \quad \epsilon_m = \frac{G_{t,m} G_{r,m} \lambda_m^2 }{(4 \pi)^3 }.
\end{equation}
In the above formula, $\beta_r$ models the amplitude and phase changes caused by the reflection at the target, $G_{t,m}$ and $G_{r,m}$ denote the antenna gain at frequency $f_m$, and $\lambda_m = c/f_m$ denote the wavelength.}

\textcolor{black}{Then, the discrete-time signal model in the $l$-th OFDM symbol after removing CP can be obtained by the sampling $y_i(t)$ at time $t = l T_{\mathrm{tot}} + T_{\mathrm{cp}} + k \frac{T_s}{M}$ for $k = 0, \dots, M-1$.} By omitting the constant terms, the noiseless discrete-time signal model is given by \cite{sturm2011waveform}:
\begin{align}
    &y_{k,i}(l) \nonumber \\ = &\frac{1}{\sqrt{M}} \sum_{n=1}^{N} \sum_{m=0}^{M-1} \beta_{m,n,i} x_{m,n}(l) e^{j 2\pi \delta_m \Delta f (\frac{k T_s}{M} - \tau_{n,i})} e^{-j 2\pi f_c \tau_{n,i} } \nonumber \\
    = &\frac{1}{\sqrt{M}} \sum_{n=1}^{N} \sum_{m=0}^{M-1} \beta_{m,n,i} x_{m,n}(l) e^{-j \frac{2\pi f_m}{c} (r_n + r_i) } e^{j 2\pi \frac{mk}{M}},
\end{align}  
where $f_m = f_c + \delta_m \Delta f$ is the frequency of the $m$-th subcarrier. Let $\tilde{\mathbf{y}}_k(l) = [ y_{k,1}(l),\dots,y_{k,N}(l)  ]^T$ denote the vector collecting all signals received at the BS, \textcolor{black}{which can be expressed as 
\begin{equation}
    \tilde{\mathbf{y}}_k(l) = \frac{1}{\sqrt{M}} \sum_{m=0}^{M-1}  \frac{\sqrt{\epsilon_m} \beta_r}{r_0^2} \tilde{\mathbf{a}}_m(\theta, r) \tilde{\mathbf{a}}_m^T(\theta, r) \bar{\mathbf{x}}_m(l) e^{j 2 \pi \frac{mk}{M}},
\end{equation}
where $\tilde{\mathbf{a}}_m(\theta, r)$ denotes the near-field array response vector and is given by 
\begin{equation} \label{array_response}
    \tilde{\mathbf{a}}_m(\theta, r) = \left[ \frac{r_0}{r_1} e^{-j k_m r_1},\dots, \frac{r_0}{r_N}e^{-j k_m r_N} \right]^T,
\end{equation}
with $r_0 = \frac{1}{N} \sum_{n=1}^N r_n$ denoting the average distance and $k_m = 2\pi f_m/c$ denoting the wavenumber.} Then, the noisy signal received on the $m$-th subcarrier in the $l$-th OFDM symbol can be obtained by discrete Fourier transform (DFT) as follows 
\begin{align}
    \mathbf{y}_m(l) = &\frac{1}{\sqrt{M}} \sum_{k=0}^{M-1} \tilde{\mathbf{y}}_k(l) e^{-j 2 \pi \frac{m k}{M}} + \mathbf{z}_m(l) \nonumber \\
    =  &\frac{\sqrt{\epsilon_m} \beta_r}{r_0^2} \mathbf{a}_m(\theta, r)\mathbf{a}_m^T(\theta, r) \bar{\mathbf{x}}_m(l) + \mathbf{z}_m(l),
\end{align}  
where $\mathbf{z}_m(l)$ denote the additive white Gaussian noise with each entry obeying i.i.d. $\mathcal{CN}(0, \sigma_w^2)$. 
To streamline the analysis, we reformulate the signal model as 
\begin{equation} \label{single_model}
    \mathbf{y}_m(l) = \beta \mathbf{A}_m(\theta, r) \mathbf{x}_m(l) + \mathbf{z}_m(l),
\end{equation}
where $\beta = \beta_r/r_0^2$, $\mathbf{A}_m(\theta, r) = \mathbf{a}_m(\theta, r)\mathbf{a}_m^T(\theta, r)$, and $\mathbf{x}_m(l) = \sqrt{\epsilon_m} \bar{\mathbf{x}}_m(l)$. 
To simplify the analysis, we assume that the scaled transmit power $\epsilon_m P_m$ has a constant value of $P$. Then, the covariance of the scaled transmit signal can be obtained from \eqref{isotropic_beam} as 
\begin{equation} \label{eqn_transmit_cov}
    \mathbf{R}_m = \mathbb{E}[ \mathbf{x}_m(l) \mathbf{x}_m^H(l) ] = \frac{P}{N} \mathbf{I}_N, \forall m.
\end{equation}
Aggregating $\mathbf{y}_m(l)$ over $L$ OFDM symbols yields the following overall signal matrix received at the $m$-th subcarrier: 
\begin{equation} \label{sensing_signal}
    \mathbf{Y}_m = [\mathbf{y}_m(1),\dots,\mathbf{y}_m(L)] = \beta \mathbf{A}_m(\theta, r) \mathbf{X}_m + \mathbf{Z}_m,
\end{equation}  
where $\mathbf{A}_m(\theta, r) = \mathbf{a}_m(\theta, r) \mathbf{a}_m^T(\theta, r)$, $\mathbf{X}_m = [\mathbf{x}_m(1),\dots,$ $\mathbf{x}_m(L)]$, and $\mathbf{Z}_m = [\mathbf{z}_m(1),\dots,\mathbf{z}_m(L)]$. 

\vspace{-0.25cm}
\subsection{\textcolor{black}{Field Boundary}}

\textcolor{black}{This subsection briefly examines the electromagnetic (EM) field boundaries related to an antenna array, identifying the valid scenarios for the proposed methods and analytical results. The EM field around an antenna array is typically divided into three regions: the reactive near field, the radiating near field, and the radiating far field. The reactive near field, dominated by wave components that decay faster than the inverse square law and do not contribute to radiation, is confined to a few wavelengths from the transmitter, even for infinitely large antenna arrays, as proved in \cite{10614327}. Therefore, its effect has been excluded from our previous models} 

Beyond the reactive near field lies the radiating field, where the EM waves exhibit normal radiation, with amplitude proportional to $1/z$ and phase $2\pi f z /c$, where $z$ and $f$ represent the propagation distance and wave frequency, respectively, as shown in \eqref{array_response}. This region is further divided into the radiating near and far fields, distinguished by diffraction behavior related to phase variation across the array aperture. The phase of the signal from the $n$-th antenna to the target at frequency $f$ is expressed as $e^{-j \frac{2\pi f}{c} r_n}$, where the propagation distance is 
\begin{equation} \label{near-field_distance}
    r_n = \| \mathbf{r} - \mathbf{q}_n \| = \sqrt{ r^2 - 2 r \mathbf{k}^T(\theta) \mathbf{q}_n + \|\mathbf{q}_n\|^2 },
\end{equation}   
with $\mathbf{k}(\theta) = [\cos \theta, \sin \theta]^T$. This precise distance must be used to characterize diffraction in the radiating near field. In the radiating far field, known as Fraunhofer diffraction, the distance $r_n$ can be approximated as:
\begin{equation} \label{far-field_distance}
    r_n \approx r - \mathbf{k}^T(\theta) \mathbf{q}_n,
\end{equation} 
derived from the first-order Taylor expansion $\sqrt{1 + x} \approx 1 + \frac{1}{2}x$. To ensure that the phase error introduced by this approximation does not exceed $\pi/8$, the condition $r \geq \frac{2D^2}{\lambda}$, where $\lambda = c/f$, must be satisfied. This defines the Rayleigh distance, marking the boundary between the radiating near and far fields. However, exact equality in \eqref{far-field_distance} is only achieved as $r \to \infty$. Therefore, to fully capture near-field effects, we use the precise distance expression in \eqref{near-field_distance} to assess phase variation across the entire radiating field.

\textcolor{black}{For clarity, we temporarily omit the amplitude variation in \eqref{array_response} by assuming $r_0 \approx r_1 \approx r_2 \approx \dots \approx r_N$. This assumption holds for ULAs when the array aperture is not excessively larger than the target distance \cite{liu2023near_tutorial}. However, for UCAs, based on \eqref{distance_UCA}, this assumption is invalid only when the radius $R$ is comparable to the target distance $r$. Under this assumption, we have $r_0/r_n \approx 1, \forall n$, allowing us to approximate the array response vector as
\begin{equation} \label{vector_approx}
    \tilde{\mathbf{a}}_m(\theta, r) \approx \mathbf{a}_m(\theta, r) = \left[ e^{-j k_m r_1}, \dots, e^{-j k_m r_N} \right]^T,
\end{equation} 
which includes only the phase variations across the antenna array.} The approximated array response vector $\mathbf{a}_m(\theta, r)$ is used throughout the paper unless otherwise noted. It is important to highlight that the spatially white transmit covariance in \eqref{eqn_transmit_cov} is optimal only for the approximated phase-only array response vector $\mathbf{a}_m(\theta, r)$. Using it with the accurate array response vector $\tilde{\mathbf{a}}_m(\theta, r)$ may lead to significant performance degradation, as will be demonstrated in Section \ref{sec:results}.

\subsection{Problem Statement}
In the considered mono-static sensing setup, the scaled data signal $\mathbf{X}_m$ is known at the BS. Therefore, the problem for near-field sensing is to estimate the remaining unknown parameters, i.e., complex channel gain $\beta$, angle $\theta$, and distance $r$ related to the target from the receive signal matrices $\{\mathbf{Y}_m\}_{m=0}^{M-1}$. Note that although $\beta$ depends on $r$, estimating $r$ from $\beta$ is challenging due to the unknown reflection coefficient $\beta_r$. Consequently, we treat $\beta$ as a single unknown parameter. In the remaining part, we focus primarily on the performance of estimating $\theta$ and $r$, which together determine the target's location.

\section{Maximum Likelihood Estimation} \label{sec:MLE}

In this section, we study the method for estimating the location parameters $\theta$ and $r$ from the received signals. From a theoretical standpoint, the minimum-variance unbiased estimator (MVUE) provides the smallest possible variance and can achieve the CRBs. However, in many practical cases, obtaining the MVUE is difficult or impossible due to the complexity of the signal model \cite[Section IV.D]{poor2013introduction}. As an alternative, maximum likelihood estimation (MLE) is commonly used and offers strong estimation performance \cite[Section IV.D]{poor2013introduction}. MLE also has good asymptotic properties, with its estimation variance approaching the CRB at high signal-to-noise ratios (SNR) and with a large number of snapshots (the number of OFDM symbols in this study).  We refer to \cite{viberg1991sensor, paulraj199316, 17564} for further discussions on the estimation variance of MLE.
%
%

We now derive the MLE of $\theta$ and $r$. In the considered conditional model, the measurement vector has a complex Gaussian distribution of $\mathbf{y}_m(l) \sim \mathcal{CN} \big( \beta \mathbf{A}_m(\theta, r) \mathbf{x}_m(l), \sigma_w^2 \mathbf{I}_{N} \big)$. Therefore, the log-likelihood function for estimating the unknown variables $\boldsymbol{\xi} = [\beta, \theta, r]^T$ from the data $\mathbf{Y}$ is given by
\begin{align}
    &\log f_{\mathbf{Y}}(\mathbf{Y}; \boldsymbol{\xi})  = - L N M \log(\pi \sigma_w^2) \nonumber \\ 
    &\hspace{1cm} - \frac{1}{\sigma_w^2} \sum_{l=0}^{L-1} \sum_{m=0}^{M-1} \left\| \mathbf{y}_m(l) - \beta \mathbf{A}_m(\theta, r) \mathbf{x}_m(l) \right\|^2.
\end{align}
Therefore, the estimator of $\boldsymbol{\xi}$ that maximize the likelihood is
\begin{align} \label{ML_problem}
    \{ \hat{\beta}, \hat{\theta}, \hat{r} \} &= \argmin_{\beta, \theta, r}\sum_{l=0}^{L-1} \sum_{m=0}^{M-1} \left\| \mathbf{y}_m(l) - \beta \mathbf{A}_m(\theta, r) \mathbf{x}_m(l) \right\|^2 \nonumber \\
    &= \argmin_{\beta, \theta, r} \sum_{m=0}^{M-1} \left\|\mathbf{Y}_m - \beta \mathbf{A}_m(\theta, r) \mathbf{X}_m  \right\|_F^2.
\end{align}
\textcolor{black}{For any given $\theta$ and $r$, the optimal estimator can be obtained as a function of $\theta$ and $r$ as follows:
\begin{align}
    \hat{\beta} = &\argmin_{\beta} \sum_{m=0}^{M-1} \left\|\mathbf{Y}_m - \beta \mathbf{A}_m(\theta, r) \mathbf{X}_m  \right\|_F^2 \nonumber \\
    = & \frac{\sum_{m=0}^{M-1} \mathrm{tr} \left( \mathbf{Y}_m \mathbf{X}^H_m \mathbf{A}^H_m(\theta, r) \right) }{ \sum_{m=0}^{M-1} \| \mathbf{A}_m(\theta, r) \mathbf{X}_m \|_F^2 }.
\end{align}    
Substituting the above solution back into \eqref{ML_problem} yields 
\begin{align}
    &\sum_{m=0}^{M-1} \left\|\mathbf{Y}_m - \hat{\beta} \mathbf{A}_m(\theta, r) \mathbf{X}_m  \right\|_F^2 \nonumber \\
    & = \sum_{m=0}^{M-1} \|\mathbf{Y}_m\|_F^2 - \frac{\left| \sum_{m=0}^{M-1} \mathrm{tr} \left( \mathbf{Y}_m \mathbf{X}^H_m \mathbf{A}^H_m(\theta, r) \right) \right|^2 }{ \sum_{m=0}^{M-1} \| \mathbf{A}_m(\theta, r) \mathbf{X}_m \|_F^2 }.
\end{align}
Since the first term $\sum_{m=0}^{M-1}\|\mathbf{Y}_m\|_F^2$ is a constant, we get the equivalent estimators of $\theta$ and $r$ as follows:
\begin{equation} \label{ML_problem_2}
    \{\hat{\theta}, \hat{r}\} = \argmax_{\theta, r} \frac{\left| \sum_{m=0}^{M-1} \mathrm{tr} \left( \mathbf{Y}_m \mathbf{X}^H_m \mathbf{A}^H_m(\theta, r) \right) \right|^2 }{ \sum_{m=0}^{M-1} \| \mathbf{A}_m(\theta, r) \mathbf{X}_m \|_F^2 }.
\end{equation}
The solutions $\hat{\theta}$ and $\hat{r}$ to the above problem can be obtained through a two-dimensional grid search over the area of interest.} However, the accuracy of the solution is limited by the grid resolution. Ensuring optimality requires a fine grid, which can result in unacceptable computational complexity. As a remedy, a coarse grid can be used initially to find an approximate solution. \textcolor{black}{This rough solution can then serve as the initial point for a Newton-type method, such as the quasi-Newton method \cite{nocedal2006numerical}, which can achieve the optimal estimates described by problem \eqref{ML_problem_2} with fast convergence and low complexity \cite{526899}.} The accuracy of the obtained solution can be evaluated by the mean-squared error (MSE), which is given by 
\begin{equation} \label{MSE_formula}
    \mathrm{MSE}_{\theta} = \frac{1}{Q} \sum_{q=1}^Q (\theta-\hat{\theta}_q)^2, \quad \mathrm{MSE}_r = \frac{1}{Q} \sum_{q=1}^Q (r-\hat{r}_q)^2,
\end{equation}
where $Q$ denotes the total number of experiments, and $\hat{\theta}_k$ and $\hat{r}_k$ denote the estimates of ground-truth $\theta$ and $r$ in the $k$-th experiment, respectively.   

\section{Performance Analysis} \label{sec:analysis}
In this section, we analyze the performance of angle and distance estimation. As deriving closed-form expressions for the MSEs in \eqref{MSE_formula} is challenging, we consider the widely-used CRB, which offers a tight lower bound on the MSEs for unbiased estimators under general conditions \cite{el2010conditional, 9439203, 6362262, wang2023cram}.
%
%
%
To derive the CRBs, we first stack the signals $\{\mathbf{Y}_m\}_{m=0}^{M-1}$ into a single vector as
\begin{align}
    \mathbf{y} = \beta \underbrace{\begin{bmatrix}
        \mathrm{vec}(\mathbf{A}_0(\theta, r) \mathbf{X}_0) \\
        \vdots \\
        \mathrm{vec}(\mathbf{A}_{M-1}(\theta, r) \mathbf{X}_{M-1})
    \end{bmatrix}}_{\mathbf{u}(\theta, r)} + \begin{bmatrix}
        \mathrm{vec}(\mathbf{Z}_0) \\
        \vdots \\
        \mathrm{vec}(\mathbf{Z}_{M-1})
    \end{bmatrix},
\end{align}
It is clear that the observation signal is complex white Gaussian, i.e., $\mathbf{y} \sim \mathcal{CN}(\beta \mathbf{u}(\theta, r), \sigma_w^2 \mathbf{I}_{NML})$. \textcolor{black}{Let $\boldsymbol{\eta} = [\theta, r, \Re(\beta), \Im(\beta)]^T$ denote the real-valued parameters to estimate, and $\mathbf{J}_{\boldsymbol{\eta}} \in \mathbb{C}^{4 \times 4}$ denotes the Fisher information matrix (FIM) for estimating $\boldsymbol{\eta}$ from $\mathbf{y}$. The entry at the $i$-row and $j$-th column of $\mathbf{J}_{\boldsymbol{\eta}}$ is given by   
\begin{equation}
    \left[ \mathbf{J}_{\boldsymbol{\eta}} \right]_{i,j} = \frac{2}{\sigma_w^2} \Re \left\{  \left(\frac{\partial \left(\beta \mathbf{u}(\theta, r)\right)}{\partial \eta_i}\right)^H \frac{\partial \left(\beta \mathbf{u}(\theta, r)\right)}{\partial \eta_j} \right\},
\end{equation}       
where $\eta_i$ is the $i$-th entry of $\boldsymbol{\eta}$. Therefore, by defining $\mathbf{u}_{\theta} = \frac{\partial \mathbf{u}}{\partial \theta}$, $\mathbf{u}_{r} = \frac{\partial \mathbf{u}}{\partial r}$, the overall FIM can be obtained as 
\begin{align}
    \mathbf{J}_{\boldsymbol{\eta}} = \frac{2}{\sigma_w^2} \begin{bmatrix}
        \mathbf{J}_{11} & \mathbf{J}_{12} \\
        \mathbf{J}_{12}^T & \mathbf{J}_{22}
    \end{bmatrix},
\end{align}
where 
\begin{align}
    \mathbf{J}_{11} = &\begin{bmatrix}
        |\beta|^2  \|\mathbf{u}_{\theta}\|^2 & |\beta|^2  \Re(\mathbf{u}_{\theta}^H \mathbf{u}_r) \\
        |\beta|^2 \Re(\mathbf{u}_{\theta}^H \mathbf{u}_r) &  |\beta|^2 \|\mathbf{u}_r\|^2 
    \end{bmatrix}, \\
    \mathbf{J}_{12} = &\begin{bmatrix}
        \Re(\beta^* \mathbf{u}_{\theta}^H \mathbf{u}) & - \Im(\beta^* \mathbf{u}_{\theta}^H \mathbf{u})\\
        \Re(\beta^* \mathbf{u}_r^H \mathbf{u}) & - \Im(\beta^* \mathbf{u}_r^H \mathbf{u})
    \end{bmatrix}, \\
    \mathbf{J}_{22} = &\begin{bmatrix}
        \|\mathbf{u}\|^2 & 0 \\
        0 & \|\mathbf{u}\|^2
    \end{bmatrix},
\end{align}
The CRBs for estimating $\theta$ and $r$ can be respectively calculated by} \cite{wang2023cram}
\begin{align}
    \label{CRB_a}
    \mathrm{CRB}_{\theta} &= \left[ \mathbf{J}_{\boldsymbol{\eta}}^{-1} \right]_{1,1} = \frac{\sigma_w^2  \|\mathbf{u}_{r} \|^2 \sin^2 \Theta}{2 |\beta|^2 \det \mathbf{Q}}, \\ 
    \label{CRB_b}
    \mathrm{CRB}_r &= \left[ \mathbf{J}_{\boldsymbol{\eta}}^{-1} \right]_{2,2} = \frac{\sigma_w^2  \|\mathbf{u}_{\theta} \|^2 \sin^2 \Omega}{2 |\beta|^2 \det \mathbf{Q}},
\end{align} 
which satisfy $\mathrm{MSE}_{\theta} \ge \mathrm{CRB}_{\theta}$ and $\mathrm{MSE}_r \ge \mathrm{CRB}_r$ for unbiased estimators.
More particularly, we have $\sin^2 \Omega = 1 - \frac{|\mathbf{u}_{\theta}^H \mathbf{u}|^2}{ \|\mathbf{u}_{\theta}\|^2 \|\mathbf{u}\|^2 }$, $\sin^2 \Theta = 1 - \frac{|\mathbf{u}_{r}^H \mathbf{u}|^2}{ \|\mathbf{u}_{r}\|^2 \|\mathbf{u}\|^2 }$, and 
\begin{equation} \label{Q_matrix}
    \mathbf{Q} = \begin{bmatrix}
        \|\mathbf{u}_{\theta} \|^2 \sin^2 \Omega & \frac{\Re( \mathbf{u}^H \mathbf{\Phi} \mathbf{u} )}{\|\mathbf{u}\|^2} \\
        \frac{\Re( \mathbf{u}^H \mathbf{\Phi} \mathbf{u} )}{\|\mathbf{u}\|^2} &  \|\mathbf{u}_{r} \|^2 \sin^2 \Theta
    \end{bmatrix},
\end{equation}
where $\mathbf{\Phi} = \mathbf{u}_{\theta}^H \mathbf{u}_r \mathbf{I} - \mathbf{u}_{\theta} \mathbf{u}_r^H$. According to \eqref{CRB_a} and \eqref{CRB_b}, the value of CRBs is determined by the terms $\|\mathbf{u}\|^2$, $\|\mathbf{u}_{\theta}\|^2$, $\|\mathbf{u}_{r}\|^2$, $\mathbf{u}_{\theta}^H \mathbf{u}_r$, $\mathbf{u}_{\theta}^H \mathbf{u}$, and $\mathbf{u}_{r}^H \mathbf{u}$. In the following, we derive these terms subject to the phase-only model \eqref{vector_approx}.

Given that $\mathbf{R}_m = \frac{P}{N} \mathbf{I}_N$, the expression of $\|\mathbf{u}\|^2$ can be derived as 
\begin{align} \label{eqn_calculate_u}
    &\|\mathbf{u}\|^2 = \sum_{m=0}^{M-1} \|\mathrm{vec}(\mathbf{A}_m(\theta, r) \mathbf{X}_m)\|^2 \nonumber \\ 
    &\overset{(a)}{\approx}  \sum_{m=0}^{M-1} \frac{ P L}{N} \mathrm{tr} \left( \mathbf{A}_m(\theta, r) \mathbf{A}^H_m(\theta, r) \right) =  P LNM,
\end{align}  
where $(a)$ stems from the equality $\|\mathrm{vec}(\mathbf{X})\|^2 = \mathrm{tr}(\mathbf{X} \mathbf{X}^H)$ and the approximation $\frac{1}{L} \mathbf{X}_m \mathbf{X}_m^H$ $\approx \mathbb{E}[ \mathbf{x}_m(l) \mathbf{x}_m(l)^H ] = \mathbf{R}_m = \frac{P}{N}\mathbf{I}_N$. 
Furthermore, to calculate the intermediate parameters involving $\mathbf{u}_{\theta}$ and $\mathbf{u}_r$, we first define the following derivatives:
\begin{align}
    \dot{\mathbf{G}}_{\theta,m} \triangleq &\frac{\partial \mathbf{A}_m(\theta, r)}{\partial \theta} = -j k_m (\dot{\mathbf{D}}_{\theta} \mathbf{A}_m(\theta, r) + \mathbf{A}_m(\theta, r) \dot{\mathbf{D}}_{\theta}), \nonumber \\
    \dot{\mathbf{G}}_{r,m} \triangleq &\frac{\partial \mathbf{A}_m(\theta, r)}{\partial r} = -j k_m (\dot{\mathbf{D}}_r \mathbf{A}_m(\theta, r) + \mathbf{A}_m(\theta, r) \dot{\mathbf{D}}_r),
\end{align}
where $\dot{\mathbf{D}}_{\theta}$ and $\dot{\mathbf{D}}_r$ are diagonal matrices whose $n$-th diagonal entries are $[\dot{\mathbf{D}}_{\theta}]_{n,n} = \frac{\partial r_n}{\partial \theta}$ and $[\dot{\mathbf{D}}_r]_{n,n} = \frac{\partial r_n}{\partial r}$, respectively.   
Then, $\mathbf{u}_{\theta}$ and $\mathbf{u}_r$ can be reformulated as 
\begin{equation}
    \mathbf{u}_i = \begin{bmatrix}
        \mathrm{vec}(\dot{\mathbf{G}}_{i,0} \mathbf{X}_0) \\
        \vdots \\
        \mathrm{vec}(\dot{\mathbf{G}}_{i,M-1} \mathbf{X}_{M-1})
    \end{bmatrix}, \forall i \in \{\theta, r\}.
\end{equation}  
Then, following the similar process as \eqref{eqn_calculate_u}, the expressions of $\|\mathbf{u}_{\theta}\|^2$ is given by 
\begin{align} \label{eq_22}
    \|\mathbf{u}_{\theta}\|^2 = &\sum_{m=0}^{M-1}  \frac{ k_m^2 P L}{N} \left\|\dot{\mathbf{D}}_{\theta} \mathbf{A}_m(\theta, r) + \mathbf{A}_m(\theta, r) \dot{\mathbf{D}}_{\theta} \right\|_F^2 \nonumber \\ = & \frac{2 k_0^2 P L M_2}{N} \left( N u_{\theta} + c_{\theta}^2 \right),
\end{align} 
where $k_0 = 2\pi/c$, $M_2 = \sum_{m=0}^{M-1} f_m^2 = M f_c^2 + \frac{M(M^2-1)}{12} \Delta f^2$, $u_{\theta} = \sum_{n=1}^N \left( \frac{\partial r_n}{\partial \theta} \right)^2$, and $c_{\theta} = \sum_{n=1}^N \frac{\partial r_n}{\partial \theta}$. 
Similarly, it can be shown that 
\begin{align}
    \label{eq_77}
    &\|\mathbf{u}_r\|^2 = \frac{2 k_0^2 P L M_2}{N} \left( N u_r + c_r^2 \right), \\
    &\mathbf{u}_{\theta}^H \mathbf{u}_r  = \frac{2 k_0^2 P L M_2}{N} \left( N \varepsilon + c_{\theta} c_r \right),\\
    &\mathbf{u}_{\theta}^H \mathbf{u} = -2 j k_0 P L M_1 c_{\theta}, \\  
    \label{eq_80}
    &\mathbf{u}_r^H \mathbf{u} = -2 j k_0 P L M_1 c_r,
\end{align}
where $u_r = \sum_{n=1}^N \left( \frac{\partial r_n}{\partial r} \right)^2$, $c_r = \sum_{n=1}^N \frac{\partial r_n}{\partial r}$, $\varepsilon = \sum_{n=1}^N  \frac{\partial r_n}{\partial \theta} \frac{\partial r_n}{\partial r}$, and $M_1 = \sum_{m=0}^{M-1} f_m = M f_c$. 

By substituting \eqref{eqn_calculate_u} and \eqref{eq_22}-\eqref{eq_80} into \eqref{CRB_a} and \eqref{CRB_b}, the CRBs for the phase-only model \eqref{vector_approx} can be obtained by
\begin{align}
    \mathrm{CRB}_{\theta} &= \frac{ N M M_2 u_{r} + (M M_2 - 2M_1^2) c_{r}^2 }{ 4 \rho L M_2 \big( N M M_2\phi + (M M_2 - 2 M_1^2) \psi\big) }, \label{CRB_theta} \\
    \mathrm{CRB}_{r} &=  \frac{ N M M_2 u_{\theta} + (M M_2 - 2M_1^2) c_{\theta}^2 }{ 4 \rho L M_2 \big( N M M_2\phi + (M M_2 - 2 M_1^2) \psi\big)}, \label{CRB_r}
\end{align}
where $\rho = k_0^2 |\beta|^2 P/\sigma_w^2$, $\phi = u_{\theta} u_r - \varepsilon^2$, and $\psi = u_{\theta} c_r^2 + u_r c_{\theta}^2 - 2 \varepsilon c_{\theta} c_r$. However, the expression of CRBs is still complicated, making it difficult to obtain useful insights. Therefore, in the following, we analyze the behavior of CRBs under different approximations in different scenarios for ULAs and UCAs, respectively.

\vspace{-0.5cm}
\textcolor{black}{
\begin{remark}
    \normalfont
    In \eqref{CRB_theta} and \eqref{CRB_r}, we derive the CRBs for the phase-only model \eqref{vector_approx}. Similarly, the CRBs for the accurate model \eqref{array_response}, which accounts for both phase and amplitude variations, can be obtained using the general CRB expressions in \eqref{CRB_a} and \eqref{CRB_b}, but can lead to a complicated form of the CRBs. Therefore, we focus on the CRBs for the phase-only model \eqref{vector_approx} to gain clearer insights. Additionally, evaluating the misspecified CRBs is an interesting direction for future work. The misspecified CRB captures the fundamental limits of estimation error when an estimator designed for the phase-only model is applied to the accurate model, which includes the classical CRB as a special case. A comprehensive derivation and review of misspecified CRBs can be found in \cite{richmond2015parameter, fortunati2017performance}.
\end{remark}
}

\vspace{-0.3cm}
\subsection{ULAs}
For ULAs, the aperture is $D = (N-1)d \approx Nd$. 
The following theorems and corollaries present an analysis of the CRBs for ULAs, focusing on the impact of the number of antennas, system bandwidth, and target distance.

\begin{figure*}[t!]
	\centering
	\begin{subfigure}{0.45\textwidth}
		\centering
		\includegraphics[width=1\textwidth]{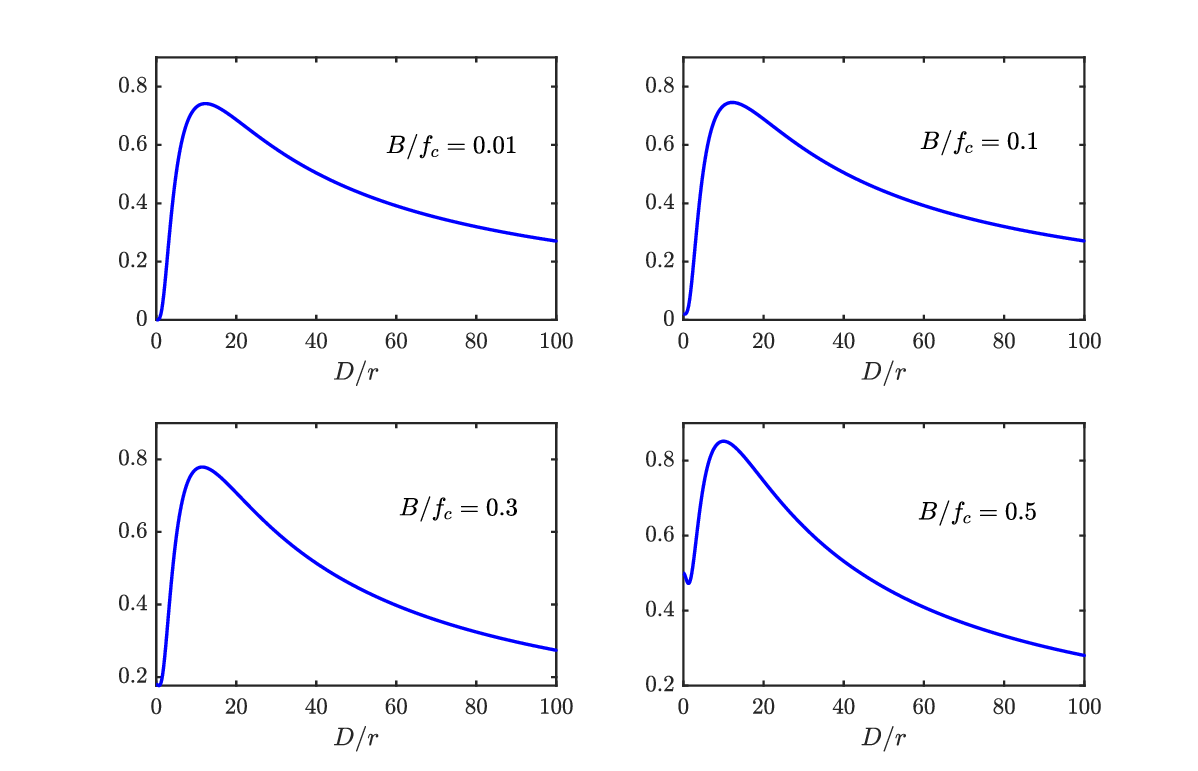}
		\caption{$0<D/r\le 100$.}
        \label{fig_ULA_au_r_full}
	\end{subfigure}
	\begin{subfigure}{0.45\textwidth}
		\centering
		\includegraphics[width=1\textwidth]{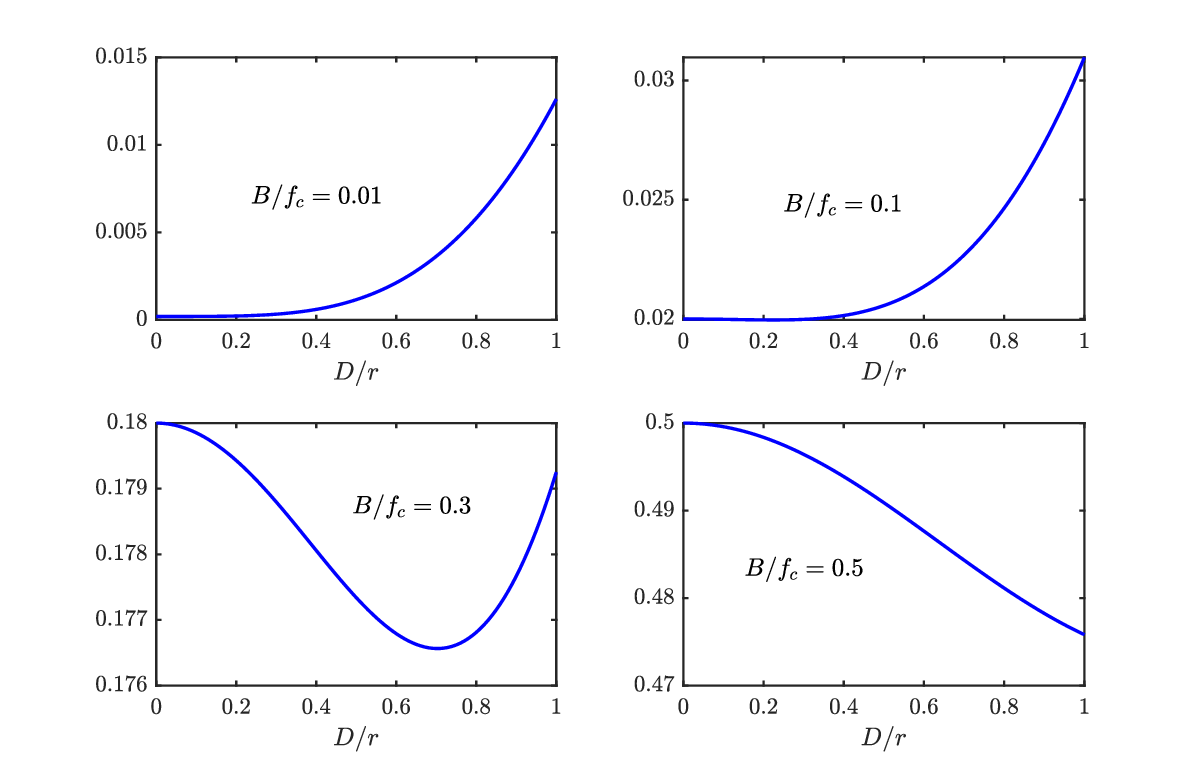}
		\caption{$0<D/r\le 1$.}
        \label{fig_ULA_au_r_partial}
	\end{subfigure}
 \caption{The numerical results of function $\Xi_r\left( \frac{D}{r}, \frac{B}{f_c} \right)$.}
 \label{fig_ULA_au_r}
\end{figure*}

\begin{theorem} \label{theorem_ULA_1}
    \normalfont
    \textcolor{black}{When $N \gg 1$, the closed-form expressions of $u_{\theta}$, $u_r$, $c_\theta$, $c_r$, and $\varepsilon$ can be derived as       
    \begin{align}
        \label{u_t_value}
        &u_{\theta} = \frac{r^3 N \sin^2 \theta}{D} \left( \frac{D}{r} + \cos \theta \ln \left( \frac{G_1}{G_2} \right) + \frac{\cos 2 \theta}{\sin \theta} \Xi  \right), \\
        \label{u_r_value}
        &u_r = N - \frac{1}{r^2} u_{\theta}, \\
        \label{c_t_value}
        &c_{\theta} = \frac{r^2 N \sin \theta}{D} \Bigg( \frac{1}{r} \left(\sqrt{ G_1 } - \sqrt{ G_2 } \right) \nonumber \\
        & \hspace{2.1cm}+ \cos \theta \ln \left(\frac{\sqrt{G_1} + \frac{1}{2}D - r \cos \theta}{ \sqrt{G_2} - \frac{1}{2}D - r \cos \theta} \right) \Bigg), \\
        \label{c_r_value}
        &c_r = \frac{r N}{D} \ln \left(\frac{\sqrt{G_1} + \frac{1}{2}D - r \cos \theta}{ \sqrt{G_2} - \frac{1}{2}D- r \cos \theta} \right) - \frac{\cos \theta}{r \sin \theta} c_{\theta}, \\
        \label{eta_value}
        &\varepsilon = \frac{r^2 N \sin \theta}{D} \left( \frac{1}{2} \ln \left( \frac{G_1}{G_2} \right) + \frac{\cos \theta}{\sin \theta} \Xi \right) - \frac{\cos \theta}{r \sin \theta} u_{\theta},
    \end{align}
    where
    \begin{align} \label{ULA_au_value}
        G_1 = &\frac{1}{4} D^2 - r D \cos \theta + r^2,
        G_2 = \frac{1}{4} D^2 + r D \cos \theta + r^2, \nonumber \\
        \Xi = &\arctan \left( \frac{D - 2 r \cos \theta}{2 r \sin \theta} \right) + \arctan \left( \frac{D + 2 r \cos \theta}{2 r \sin \theta} \right).
    \end{align} }
\end{theorem}
\vspace{-0.1cm}
\begin{IEEEproof}
    Please refer to Appendix \ref{proof_theorem_ULA_1}.
\end{IEEEproof}



By substituting \eqref{u_t_value}-\eqref{eta_value} into \eqref{CRB_theta}, and \eqref{CRB_r}, the closed-form expression of CRBs in \eqref{CRB_a} and \eqref{CRB_b} can be calculated. However, the resulting expressions are still too complex to obtain useful insights. Therefore, in the following theorem, we investigate the CRBs when the target is along the broadside of the ULA, i.e., $\theta = \pi/2$, so that the maximum effective aperture is attained.

\begin{corollary} \label{Theorem_ULA_broadside}
    \normalfont
    \textcolor{black}{When $\theta = \pi/2$ and $N \gg 1$, the closed-form expression of CRBs are given by  
    \begin{align}
        \label{CRB_theta_ULA_approx}
        \mathrm{CRB}_{\theta} &= \frac{3}{\rho L N M r^2 \Phi\left( \frac{D}{r} \right) (12 f_c^2 + B^2 - \Delta f^2) }, \\
        \mathrm{CRB}_r &= \frac{3}{\left( \splitdfrac{\rho L N M \Big[ 12 f_c^2 \big( 1 - \Phi\left(\frac{D}{r}\right) - \Psi^2\left(\frac{D}{r}\right)\big)}{ + (B^2 - \Delta f^2) \left( 1 - \Phi\left(\frac{D}{r}\right) + \Psi^2\left(\frac{D}{r}\right) \right)  \Big]} \right)},
    \end{align}
    Functions $\Phi(\alpha)$ and $\Psi(\alpha)$ are given by 
    \begin{align}
        &\Phi(\alpha) = 1 - \frac{2}{\alpha} \mathrm{arctan} \left( \frac{\alpha}{2}\right), \\
        &\Psi(\alpha) = \frac{1}{\alpha} \ln \left( \frac{\sqrt{\alpha^2 + 4} + \alpha}{\sqrt{\alpha^2 + 4} - \alpha} \right).
    \end{align}}
\end{corollary}

\begin{IEEEproof}
    This corollary can be readily proved by substituting $\theta = \pi/2$ into the results in \textbf{Theorem \ref{theorem_ULA_1}} and CRBs in \eqref{CRB_theta} and \eqref{CRB_r}.
\end{IEEEproof}
 

\begin{figure}[t!]
    \centering
    \includegraphics[width=0.45\textwidth]{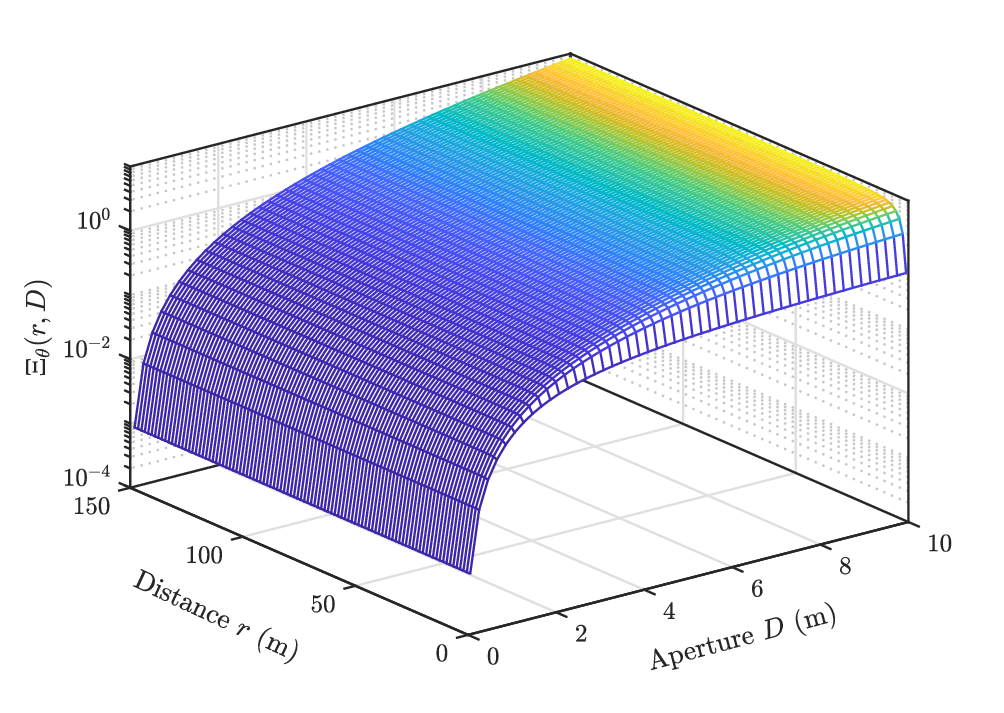}
    \caption{The numerical results of function $\Xi_{\theta}(r, D)$.}
    \label{fig_ULA_au_theta}
\end{figure}

In \textbf{Corollary \ref{Theorem_ULA_broadside}}, it can be observed that $\mathrm{CRB}_{\theta}$ and $\mathrm{CRB}_r$ are inversely proportional to the following functions, respectively:
\begin{align}
    &\Xi_{\theta}(r, D) =  r^2 \Phi\left( \frac{D}{r} \right), \nonumber \\
    &\Xi_r\left( \frac{D}{r}, \frac{B}{f_c} \right) = 12 \left( 1 - \Phi\left(\frac{D}{r}\right) - \Psi^2\left(\frac{D}{r}\right)\right) \nonumber \\
    &\hspace{1cm}+ \frac{B^2 - \Delta f^2}{f_c^2} \left( 1 - \Phi\left(\frac{D}{r}\right) + \Psi^2\left(\frac{D}{r}\right) \right).
\end{align}
The behaviors of these function when $M \gg 1$, i.e., $B^2 - \Delta f^2 \approx B^2$, are illustrated in Fig. \ref{fig_ULA_au_theta} and Fig. \ref{fig_ULA_au_r}, respectively. From these numerical results and \textbf{Corollary \ref{Theorem_ULA_broadside}}, we notice the following insights for a target located along the broadside direction, i.e., $\theta = \pi/2$, of the ULA.

\begin{remark} \label{remark_1}
    \normalfont
    \textcolor{black}{\emph{(Impact of Bandwidth)} Both $\mathrm{CRB}_{\theta}$ and $\mathrm{CRB}_r$ are $O(1/M)$ and decrease with both larger carrier frequency $f_c$ and bandwidth $B$. Compared to $\mathrm{CRB}_r$, $\mathrm{CRB}_{\theta}$ is less affected by the bandwidth $B$. This is because the $(12 f_c^2 + B^2 - \Delta f^2)$ term in its denominator is mainly affected by $12 f_c^2$ unless $B^2 - \Delta f^2$ has a comparable value to $12f_c^2$, which is typically impossible in practice. In contrast, $\mathrm{CRB}_r$ is significantly affected by the bandwidth $B$, which will be detailed in the following remark.}
\end{remark}

\begin{remark} \label{remark_2}
    \normalfont
    \textcolor{black}{\emph{(Impact of Aperture Size)} $\mathrm{CRB}_{\theta}$ is $O(1/N)$ and is dependent on both distance $r$ and aperture $D$, as described by the function $\Xi_{\theta}(r, D)$. Fig. \ref{fig_ULA_au_theta} indicates that the improvements in angle estimation from increasing the aperture \(D\) gradually diminish as $D$ becomes larger. This phenomenon is further illustrated by the limit $\lim_{D \rightarrow +\infty} \Phi(\frac{D}{r}) = 1$. Additionally, Fig. \ref{fig_ULA_au_theta} also suggests that $\mathrm{CRB}_{\theta}$ is independent of $r$ except when the target is very close to the ULA, where $a$ smaller $r$ decreases the performance of angle estimation. On the contrary, $\mathrm{CRB}_r$ is $O(1/N)$ but exhibits a more complex dependence $r$ and $D$, characterized by the function $\Xi_r(\frac{D}{r}, \frac{B}{f_c})$. According to Fig. \ref{fig_ULA_au_r}(a), regardless of the bandwidth, there exists an optimal radio $D/r$ that maximizes the distance sensing performance, indicating that a larger aperture $D$ or a closer target do not necessarily lead to a better sensing performance. Furthermore, for the practical case when $D/r \le 1$ as illustrated in Fig. \ref{fig_ULA_au_r}(b), larger array apertures or closer targets (i.e., a larger ratio of $D/r$) generally enhance the performance of distance estimation with the practical ratio $B/f_c$ less than $0.1$. However, this trend appears to reverse at ultra-high values of $B/f_c$, exceeding $0.1$, where larger apertures or closer targets may actually degrade the performance of distance estimation. It is important to note that such high values of $B/f_c$ are generally uncommon in practice.}
\end{remark}

\begin{remark} \label{remark_3}
    \normalfont
    \textcolor{black}{\emph{(Origins of Performance Gain)} In the considered system, the total number of observations for target sensing is $LMN$, where $L$, $M$, and $N$ denote the number of observations collected across the time, frequency, and space domains, respectively. Recall that with fixed $D$, and $B$, both $\mathrm{CRB}_{\theta}$ and $\mathrm{CRB}_r$ are $O(1/N)$ and $O(1/M)$. This suggests that when the array aperture and signal bandwidth remain constant, the performance gain from adding more antennas and subcarriers stems solely from the increase in the total number of observations, rather than changing the near-field and wideband effects. Conversely, modifications to the array aperture or signal bandwidth directly change these effects, thereby fundamentally altering sensing performance.} 
\end{remark}

In the preceding discussion, we studied the CRBs for a broadside target. Our numerical results in Section \ref{sec:results} will demonstrate that the above conclusions also apply to an arbitrarily located target. To gain further theoretical insights for an arbitrarily located target, in the following, we investigate the estimation performance through the asymptotic analysis. We first focus on the impact of array size, for which we have the following results. 

\begin{corollary} \label{Theorem_ULA_N_limit_0}
    \normalfont
    \textcolor{black}{For fixed aperture $D$, as $N \rightarrow +\infty$, the asymptotic CRBs achieved by a ULA are given by
    \begin{equation}
        \lim_{N \rightarrow +\infty} \mathrm{CRB}_{\theta} = 0, \quad \lim_{N \rightarrow +\infty} \mathrm{CRB}_r = 0.
    \end{equation}}
\end{corollary}

\begin{IEEEproof}
    The above limits can be readily obtained. We thus omit the proof here.
\end{IEEEproof} 

\begin{corollary} \label{Theorem_ULA_N_limit}
    \textcolor{black}{\normalfont For fixed antenna spacing $d = D/N$, as $N \rightarrow +\infty$, the asymptotic CRBs achieved by a ULA are given by
    \begin{align} \label{ULA_N_limit_r}
        \lim_{N \rightarrow +\infty} \mathrm{CRB}_{\theta} &= \frac{3 d \cos^2 \theta}{\rho L M  \pi r^3 (12f_c^2 + B^2 - \Delta f^2) \sin \theta}, \\
        \lim_{N \rightarrow +\infty} \mathrm{CRB}_r &= \frac{3 d \sin \theta}{\rho L M \pi r (12f_c^2 + B^2 - \Delta f^2)}.
    \end{align}}  
\end{corollary}

\begin{IEEEproof}
    Please refer to Appendix \ref{proof_Theorem_ULA_N_limit}.
\end{IEEEproof}

\begin{remark} \label{remark_ULA_1}
    \normalfont
    \textbf{Corollaries \ref{Theorem_ULA_N_limit_0}} and \textbf{\ref{Theorem_ULA_N_limit}} present the behavior of CRBs as the number of antennas $N$ increases under two conditions: 1) with a fixed aperture, and 2) with fixed antenna spacing respectively. Under the first condition, increasing $N$ will decrease CRBs unboundedly, while under the second condition, the CRBs are bounded from below. However, this does not mean that these two theorems present conflicting results. Actually, under the first condition, we have $d \rightarrow 0$ as $N \rightarrow +\infty$. Substituting $d \rightarrow 0$ into the results in \textbf{Corollary \ref{Theorem_ULA_N_limit}}, we will get the same results as in \textbf{Corollary \ref{Theorem_ULA_N_limit_0}}. \textcolor{black}{The above observation highlights the importance of reducing antenna spacing for near-field sensing, revealing the significant potential of emerging antenna array technologies with sub-half-wavelength spacing, such as dynamic metasurface antennas \cite{9324910}, or those with ideally continuous apertures \cite{liu2024capa}.}
    Additionally, the results in \textbf{Corollary~\ref{Theorem_ULA_N_limit}} show that the near-field sensing performance of a ULA depends on both the target's direction and distance.
\end{remark}

\begin{corollary} \label{theorem_ULA_D_limit}
    \normalfont 
    For a fixed number of antennas $N$, as $D \rightarrow +\infty$, the asymptotic CRBs achieved by a ULA are given by 
    \begin{align}
        &\lim_{D \rightarrow +\infty} \mathrm{CRB}_{\theta} = \begin{cases}
            \frac{3}{\rho L N M r^2 (12 f_c^2 + B^2 - \Delta f^2) }, &\text{if } \theta = \frac{\pi}{2}, \\
            +\infty, &\text{if } \theta \neq \frac{\pi}{2}.
        \end{cases} \\
        &\lim_{D \rightarrow +\infty} \mathrm{CRB}_r = +\infty.
    \end{align}
\end{corollary}

\begin{IEEEproof}
    Please refer to Appendix \ref{proof_theorem_ULA_D_limit}.
\end{IEEEproof}

\textcolor{black}{
\begin{remark} \label{remark_6}
    \normalfont 
    \textbf{Corollaries \ref{Theorem_ULA_N_limit}} and \textbf{\ref{theorem_ULA_D_limit}} both describe the behavior of CRBs as the array aperture $D$ approaches infinity, considering fixed antenna spacing and a fixed number of antennas, respectively. Although the approximation in \eqref{vector_approx} becomes invalid under these conditions, these corollaries still provide valuable insights.  In particular, the results in these two corollaries reveal that under the approximated phase-only model \eqref{vector_approx}, the CRBs either converge to a bound or diverge to infinity as $D \rightarrow +\infty$. This, in turn, implies that under the accurate model \eqref{array_response}, the CRBs must approach infinity. The key reason behind this phenomenon is the use of the isotropic beam in \eqref{isotropic_beam}, which distributes equal power across all antennas.  More specifically, for the accurate model, as $D \rightarrow +\infty$, an increasing number of antennas at the edges of the ULA become ineffective due to the significant pathloss caused by their large distance from the target. This reduces the effective transmit power available for target sensing, leading to degraded performance. Consequently, the sensing performance characterized by the accurate model must be worse than that predicted by \textbf{Corollaries \ref{Theorem_ULA_N_limit}} and \textbf{\ref{theorem_ULA_D_limit}}, ultimately resulting in unbounded CRBs. This finding further confirms that increasing the aperture $D$ does not necessarily improve sensing performance.
\end{remark}
}


\begin{corollary} \label{theorem_ULA_far}
    \normalfont
    \textcolor{black}{As $r \rightarrow +\infty$, the asymptotic CRBs achieved by a ULA are given by 
    \begin{align}
        \lim_{r \rightarrow +\infty} \mathrm{CRB}_{\theta} &= \frac{36}{ \rho L N M D^2 \left( 12 f_c^2 + B^2 - \Delta f^2 \right) \sin^2 \theta }, \\
        \lim_{r \rightarrow +\infty} \mathrm{CRB}_r &= \frac{3}{2\rho L N M \left(B^2 - \Delta f^2 \right)}.
    \end{align}}
\end{corollary}

\begin{IEEEproof}
    Please refer to Appendix \ref{proof_theorem_ULA_far}
\end{IEEEproof}

\begin{figure*}
	\centering
	\begin{subfigure}{0.45\textwidth}
		\centering
		\includegraphics[width=1\textwidth]{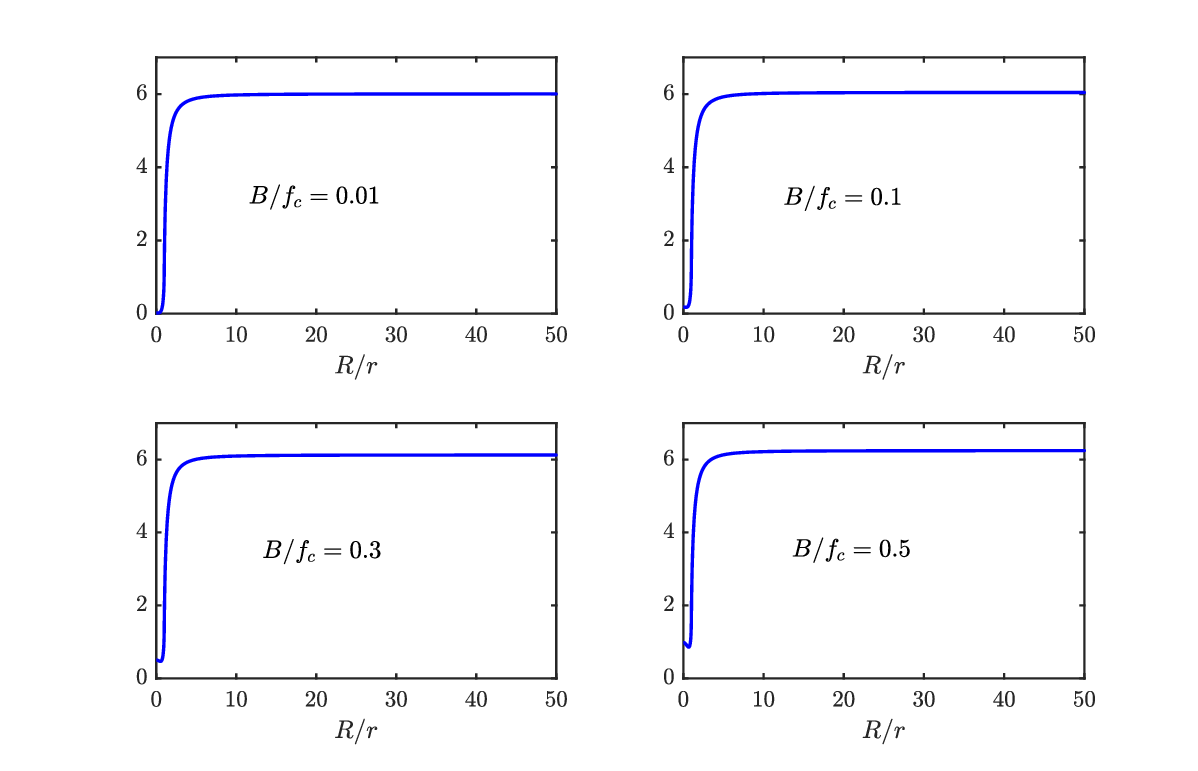}
		\caption{$0<R/r\le 50$.}
        \label{fig_UCA_au_r_full}
	\end{subfigure}
	\begin{subfigure}{0.45\textwidth}
		\centering
		\includegraphics[width=1\textwidth]{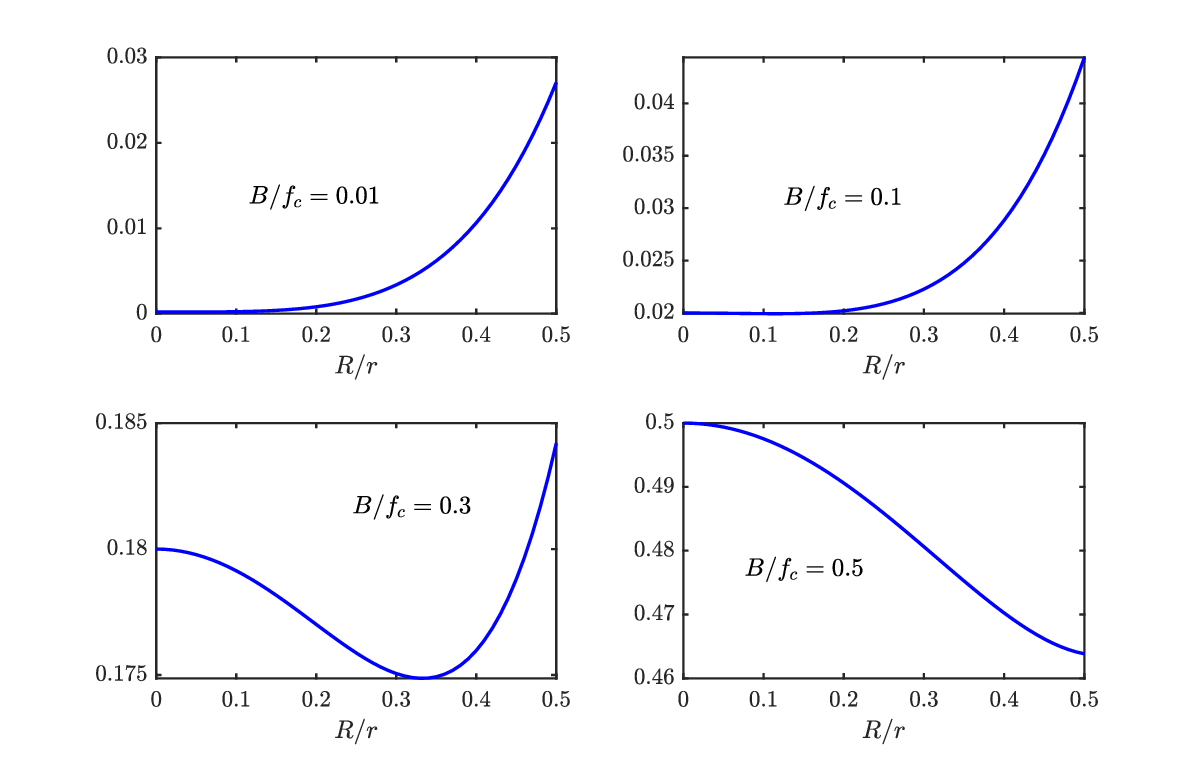}
		\caption{$0<R/r\le 0.5$}
        \label{fig_UCA_au_r_partial}
	\end{subfigure}
 \caption{The numerical results of function $\tilde{\Xi}_r\left( \frac{R}{r}, \frac{B}{f_c} \right)$. }
 \label{fig_UCA_au_r}
\end{figure*}

\begin{remark}
    \normalfont
    \textcolor{black}{\textbf{Corollary \ref{theorem_ULA_far}} describes the CRBs in far-field sensing systems (note that the far-field approximation in \eqref{far-field_distance} is achieved with exact equality only when $r \rightarrow \infty$). Under this circumstance, when the number of observations $LNM$ remains constant, $\mathrm{CRB}_r$ is irrelevant to the array aperture $D$ and is merely related to the bandwidth $B$. This finding aligns with previous research on far-field sensing \cite{guerci2015joint}. Additionally, in a single-carrier system where $M = 1$, we have $B^2 - \Delta f^2 = 0$ so that $\mathrm{CRB}_r = +\infty$, implying the infeasibility of distance in single-carrier far-field systems. Regarding $\mathrm{CRB}_{\theta}$, it is $O(1/D^2)$ in far-field systems, which is different from the near-field cases shown in \textbf{Corollary \ref{Theorem_ULA_broadside}}.}  
\end{remark}

\begin{figure*}[b]
    \vspace*{8pt}
	\hrulefill
	\vspace*{8pt}
    \begin{align} \label{K_func}
        \tilde{\Xi}_r\left(\frac{R}{r}, \frac{B}{f_c}\right) = \begin{dcases}
            12 \left( 1 - \frac{R^2}{2r^2} - \Upsilon^2 \left( \frac{r}{R} \right)  \right) + \frac{B^2 - \Delta f^2}{f_c^2} \left( 1 - \frac{R^2}{2 r^2} + \Upsilon^2 \left( \frac{r}{R} \right) \right), &\text{if } R < r \\
            12 \left( \frac{1}{2}  - \Upsilon^2 \left( \frac{r}{R} \right)  \right) + \frac{B^2 - \Delta f^2}{f_c^2} \left( \frac{1}{2}  + \Upsilon^2 \left( \frac{r}{R} \right) \right), &\text{if } R \ge r 
        \end{dcases} \tag{67}
    \end{align}
\end{figure*}

\subsection{UCAs}

For UCAs, the aperture is $D = 2R = Nd/\pi$. 
Then, we have the following theorem for CRBs achieved by a UCA.

\setcounter{equation}{61}
\begin{theorem} \label{theorem_UCA}
    \normalfont
    When $N \gg 1$, the closed-form CRBs achieved by UCAs for $R < r$ are given by 
    \begin{align} \label{cf_crb_t}
        \mathrm{CRB}_{\theta} = &\frac{6}{ \rho  L N M R^2 \left( 12 f_c^2 + B^2 - \Delta f^2  \right)}, \\
        \label{cf_crb_r}
        \mathrm{CRB}_r = &\frac{3}{\left( \splitdfrac{\rho L N M \Big[ 12 f_c^2 \big( 1 - \frac{R^2}{2r^2} - \Upsilon^2\left( \frac{r}{R} \right) \big)}{ + \left(B^2 - \Delta f^2\right) \left( 1 - \frac{R^2}{2r^2} + \Upsilon^2\left( \frac{r}{R} \right) \right) \Big] } \right)},
    \end{align}
    where $\Upsilon(\alpha)$ is a monotonically increasing transcendental function, given by 
    \begin{equation} 
        \Upsilon(\alpha) = \int_{0}^{2 \pi} \frac{\alpha - \cos x}{2\pi \sqrt{ 1 - 2 \alpha \cos x + \alpha^2 }} dx.
    \end{equation} 
    The closed-form CRBs for $R \ge r$ are given by 
    \begin{align}  \label{cf_crb_t_2}
        \mathrm{CRB}_{\theta} = &\frac{6}{ \rho  L N M r^2 \left( 12 f_c^2 + B^2 - \Delta f^2  \right)}, \\
        \label{cf_crb_r_2}
        \mathrm{CRB}_r = &\frac{3}{ \left( \splitdfrac{\rho L N M \Big[ 12 f_c^2 \left( \frac{1}{2} - \Upsilon^2\left( \frac{r}{R} \right) \right)}{ + \left(B^2 - \Delta f^2\right) \left( \frac{1}{2} + \Upsilon^2\left( \frac{r}{R} \right) \right) \Big] } \right)},
    \end{align}
\end{theorem}

\begin{IEEEproof}
    Please refer to Appendix \ref{proof_theorem_UCA}.
\end{IEEEproof}

In \textbf{Theorem \ref{theorem_UCA}}, it can be observed that $\mathrm{CRB}_r$ achieved by a UCA is inversely proportional to the function $\tilde{\Xi}_r\left(\frac{R}{r}, \frac{B}{f_c}\right)$, which is given in \eqref{K_func} at the bottom of this page.
The behavior of this function when $M \gg 1$, i.e., $B^2 - \Delta f^2 \approx B^2$ is illustrated in Fig. \ref{fig_UCA_au_r}. From these numerical results and \textbf{Theorem \ref{theorem_UCA}}, we notice the following differences between ULAs and UCAs (keeping in mind $D = 2R$). 

\begin{remark} \label{remark_7}
    \normalfont
    \emph{(Robustness of UCA)} $\mathrm{CRB}_{\theta}$ and $\mathrm{CRB}_r$ achieved by UCA is independent of the angle of the target $\theta$. This is fundamentally different from the conventional ULAs with an angular-dependent performance \cite{el2010conditional, 9439203, 6362262, wang2023cram}. Therefore, UCAs can provide more stable performance than ULAs.
\end{remark}

\begin{remark}
    \normalfont 
    \textcolor{black}{\emph{(Outside Target)} The sensing performance achieved by UCA depends on whether the target is located inside or outside the UCA. For an outside target described by \eqref{cf_crb_t} and \eqref{cf_crb_r}, $\mathrm{CRB}_{\theta}$ achieved by a UCA is $O(1/N)$ and $O(1/D^2)$ but is independent of $r$. This is fundamentally different from the scaling law for a ULA. Specifically, for UCAs with outside targets, angle estimation always benefits from increasing $D$ whereas in ULA systems, these benefits gradually diminish as $D$ becomes large. Furthermore, if the antenna spacing $d$ remains constant, the aperture $D$ is proportional to the number of antennas, where $D = Nd/\pi$. In this case, $\mathrm{CRB}_{\theta}$ achieved by UCAs is $O(1/N^3)$. This is also fundamentally different from ULA systems, where $\mathrm{CRB}_{\theta}$ is bounded from below as $N$ increases in the same case with fixed $d$, c.f. \eqref{ULA_N_limit_r}. Regarding $\mathrm{CRB}_r$, according to the results in Fig. \eqref{fig_UCA_au_r}, a similar conclusion holds as in ULA systems, i.e., a larger aperture may not always enhance the performance of distance estimation especially when $B/f_c$ is large.} 
\end{remark}

\begin{remark}
    \normalfont
    \textcolor{black}{\emph{(Inside Target)} The case with an inside target can be regarded as a sensing with a distributed array \cite{liu2022integrated}. As described by \eqref{cf_crb_t_2} and \eqref{cf_crb_r_2}, in this case, $\mathrm{CRB}_{\theta}$ becomes independent of the aperture $D$ but depends on the target distance $r$. Furthermore, a larger distance leads to a better performance of angle estimation, which is similar to the ULA systems. Regarding $\mathrm{CRB}_r$, for an inside target, it gradually converges to a stable value as the aperture $D$ increases, as shown in Fig. \ref{fig_UCA_au_r}(a).}
\end{remark}

In the following, we further study the asymptotic CRBs achieved by a UCA, which are given by the following corollaries.

\setcounter{equation}{67}
\begin{corollary} \label{corollary_UCA_1}
    \normalfont
    \textcolor{black}{For fixed aperture $D$, as $N \rightarrow +\infty$, the asymptotic CRBs achieved by a UCA are given by 
    \begin{equation}
        \lim_{N \rightarrow +\infty} \mathrm{CRB}_{\theta} = 0, \quad \lim_{N \rightarrow +\infty} \mathrm{CRB}_r = 0.
    \end{equation} }
\end{corollary}

\begin{IEEEproof}
    The above limits can be readily obtained from the results in \textbf{Theorem \ref{theorem_UCA}}. 
\end{IEEEproof}

\begin{corollary} \label{corollary_UCA_2}
    \normalfont
    \textcolor{black}{For fixed antenna spacing $d = \pi D/N$, as $N \rightarrow +\infty$, the asymptotic CRBs achieved by a UCA are given by 
    \begin{equation}
        \lim_{N \rightarrow +\infty} \mathrm{CRB}_{\theta} = 0, \quad \lim_{N \rightarrow +\infty} \mathrm{CRB}_r = 0.
    \end{equation}}
\end{corollary}

\begin{IEEEproof}
    For fixed antenna spacing $d = 2 \pi R/N$, as $N \rightarrow +\infty$, we must have $R > r$. Thus, the first limit can be readily obtained according to \eqref{cf_crb_t_2}. For the second limit, we have 
    \begin{equation}
        \lim_{N \rightarrow +\infty} \Upsilon\left( \frac{r}{R} \right) = \lim_{N \rightarrow +\infty} \Upsilon\left( \frac{2 \pi r}{Nd} \right) = 0.
    \end{equation}
    Based on the above limit and the expression in \eqref{cf_crb_r_2}, the limit of $\mathrm{CRB}_r$ can also be readily obtained. The proof is thus completed. 
\end{IEEEproof}

\begin{remark} \label{remark_UCA_1}
    \normalfont
    \textbf{Corollaries \ref{corollary_UCA_1}} and \textbf{\ref{corollary_UCA_2}} demonstrate a different asymptotic behavior of UCAs from ULAs as $N \rightarrow +\infty$. More specifically, for fixed antenna spacing $d$, the asymptotic CRBs achieved by ULAs are bounded from below, whereas those achieved by UCAs are not.
\end{remark}

\begin{corollary} \label{corollary_UCA_3}
    \normalfont 
    \textcolor{black}{For fixed number of antennas $N$, as $D \rightarrow +\infty$, the asymptotic CRBs achieved by a UCA are given by 
    \begin{align}
        \lim_{D \rightarrow +\infty} \mathrm{CRB}_{\theta} = &\frac{6}{ \rho  L N M r^2 \left( 12 f_c^2 + B^2 - \Delta f^2  \right)} \\
        \lim_{D \rightarrow +\infty} \mathrm{CRB}_r = &\frac{6}{\rho L N M\left( 12 f_c^2 + B^2 - \Delta f^2 \right)}.
    \end{align}}
\end{corollary}

\begin{IEEEproof}
    The proof resembles the proof of \textbf{Corollary} \ref{corollary_UCA_2}.
\end{IEEEproof}

\begin{remark}
    \normalfont 
    \textbf{Corollary \ref{corollary_UCA_3}} suggests that increasing only the array aperture $D$ results in the CRBs for UCAs converging to a stable value. This result is fundamentally different from the result in \textbf{Corollary \ref{theorem_ULA_D_limit}} for ULAs, where the CRBs approach infinity as $D$ grows. \textcolor{black}{It is also noteworthy that the approximation \eqref{vector_approx} is still valid for UCAs when $D \rightarrow +\infty$.}
\end{remark}

\begin{corollary} \label{theorem_UCA_limit_r}
    \normalfont
    As $r \rightarrow +\infty$, the asymptotic CRBs achieved by a UCA are given by 
    \begin{align}
        \lim_{r \rightarrow +\infty} \mathrm{CRB}_{\theta} &= \frac{6}{ \rho  L N M R^2 \left( 12 f_c^2 + B^2 - \Delta f^2  \right)}, \\
        \label{cf_crb_r_r}
        \lim_{r \rightarrow +\infty} \mathrm{CRB}_r &= \frac{3}{2\rho L N M \left(B^2 - \Delta f^2\right)}.
    \end{align} 
\end{corollary}

\begin{IEEEproof}
    The first limit related to $\mathrm{CRB}_{\theta}$ can be directly obtained from \eqref{cf_crb_t} which is independent of $r$. For the second limit related to $\mathrm{CRB}_r$, we have $\lim_{r \rightarrow \infty} \frac{R}{2 r^2} = 0$ and 
    \begin{equation}
        \lim_{r \rightarrow +\infty} \Upsilon\left( \frac{r}{R} \right) = \int_{0}^{2 \pi} \frac{1}{2\pi} dx = 1.
    \end{equation} 
    By substituting these two limits into \eqref{cf_crb_r}, the expression in \eqref{cf_crb_r_r} can be obtained.
\end{IEEEproof}

\begin{figure}[t!]
    \centering
    \includegraphics[width=0.48\textwidth]{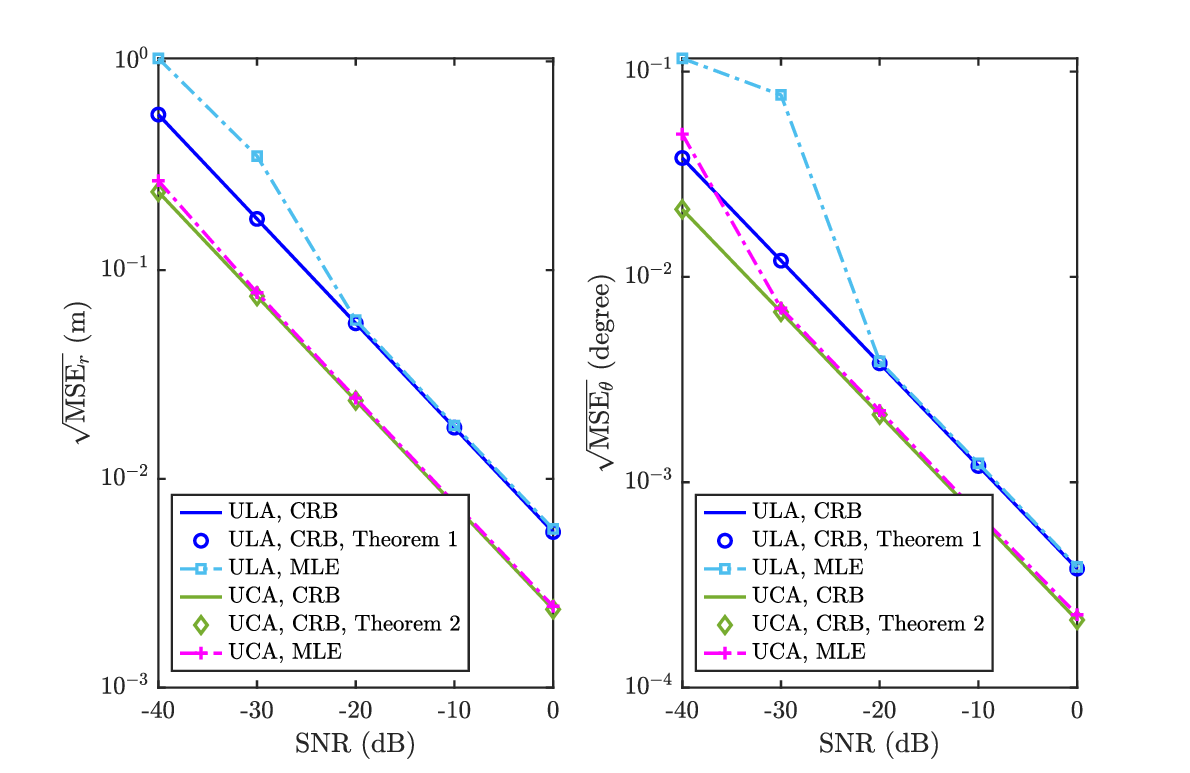}
    \caption{\textcolor{black}{Comparison between the derived CRBs and the MSEs achieved by MLE.}}
	\label{fig_MLE}
\end{figure}

\begin{remark}
    \normalfont
    \textcolor{black}{By comparing \textbf{Corollaries \ref{theorem_ULA_far}} and \textbf{\ref{theorem_UCA_limit_r}}, it can be observed that ULA and UCA exhibit the same performance of distance estimation in far-field cases, which are merely related to the bandwidth $B$. This is expected because the antenna array cannot resolve the distance information in the space domain in far-field cases. Thus, the performance of distance estimation is irrelevant to the array geometry.}
\end{remark}

\section{Numerical Results} \label{sec:results}
In this section, numerical results are provided to validate the derived CRB results. Unless otherwise specified, we set $f_c = 28$ GHz, $B = 10$ MHz, $D = 5$ m, $L = 256$, $N = 256$, $M = 256$, $|\beta|^2 = -30$ dB, $r = 20$ m, and $\theta = \pi/4$. The signal-to-noise ratio (SNR), which is defined as $\mathrm{SNR} = P/\sigma_w^2$, is set to $0$ dB.

\subsection{Performance of MLE}
\textcolor{black}{Fig. \ref{fig_MLE} evaluates the MSEs achieved by the MLE method in comparison to the derived CRBs. The MSEs are obtained by averaging $K = 500$ experiments. It can be observed that for both distance and angle estimation, the MLE method achieves MSEs comparable to the CRBs, except in cases of very low SNR (e.g., less than $-20$ dB). The approximations derived in \textbf{Theorems \ref{theorem_ULA_1}} and \textbf{\ref{theorem_UCA}} are also consistent with the true values of the CRBs. These results validate not only the effectiveness of the MLE method but also the correctness of the derived CRBs. Additionally, under the same aperture size, the UCA exhibits lower MSEs for both distance and angle estimation, confirming its advantages for sensing applications.} 

\begin{figure}[t!]
	\centering
	\begin{subfigure}{0.48\textwidth}
		\centering
		\includegraphics[width=1\textwidth]{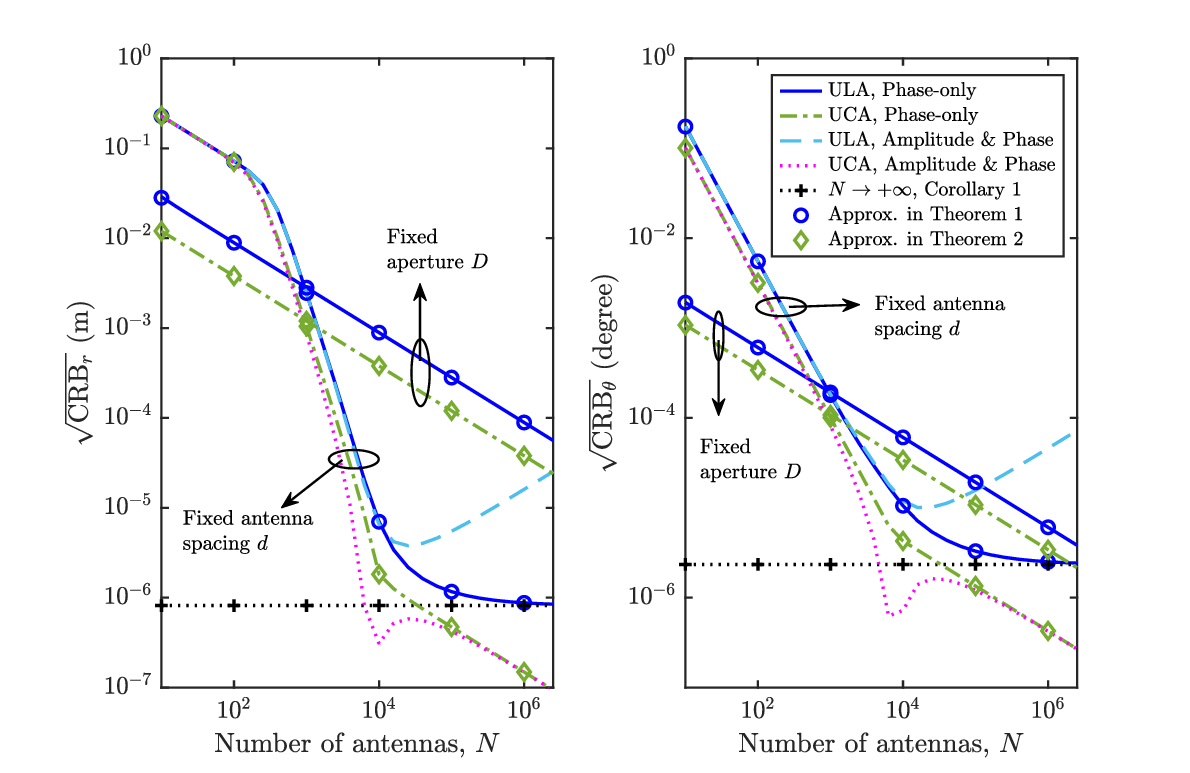}
		\caption{$\theta = \pi/4$}
	\end{subfigure}
	\begin{subfigure}{0.48\textwidth}
		\centering
		\includegraphics[width=1\textwidth]{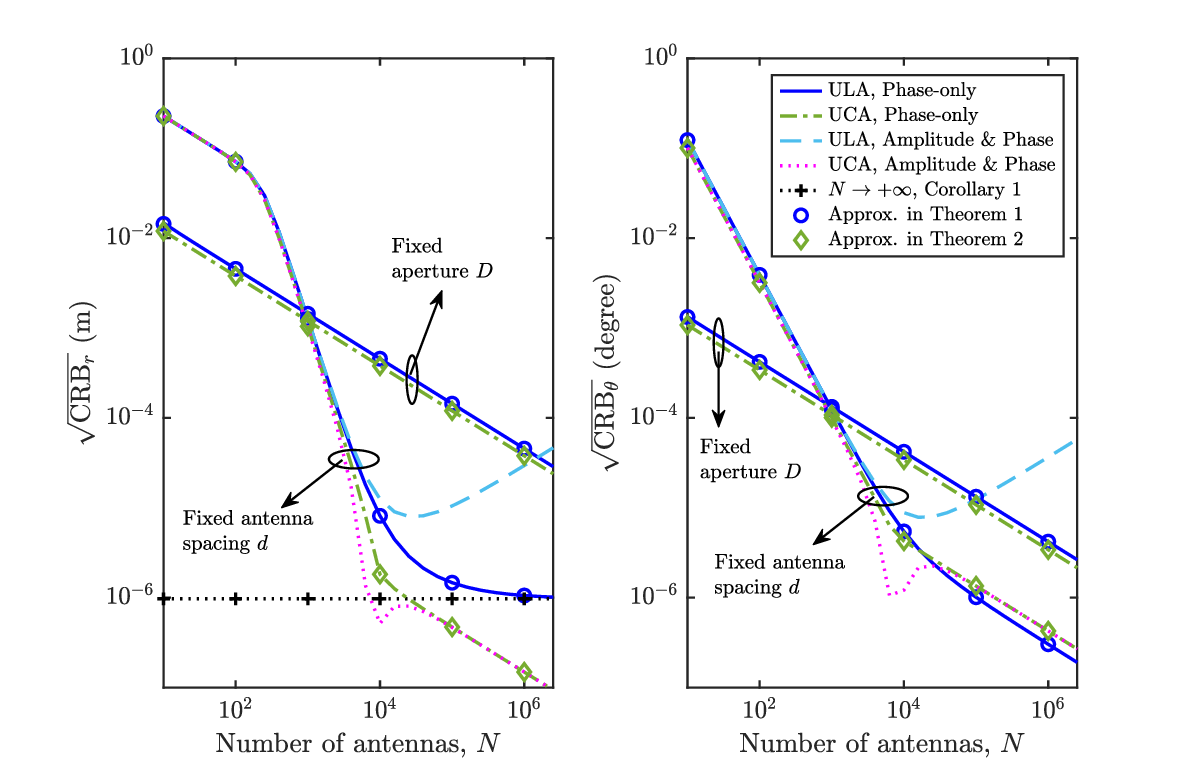}
		\caption{$\theta = \pi/2$.}
	\end{subfigure}
 \caption{\textcolor{black}{CRBs versus the number of antennas under the conditions of fixed array aperture and fixed antenna spacing, respectively.}}
 \label{fig_antenna_number}
\end{figure}

\subsection{Impact of Array Size}

\begin{figure}[t!]
	\centering
	\begin{subfigure}{0.48\textwidth}
		\centering
		\includegraphics[width=1\textwidth]{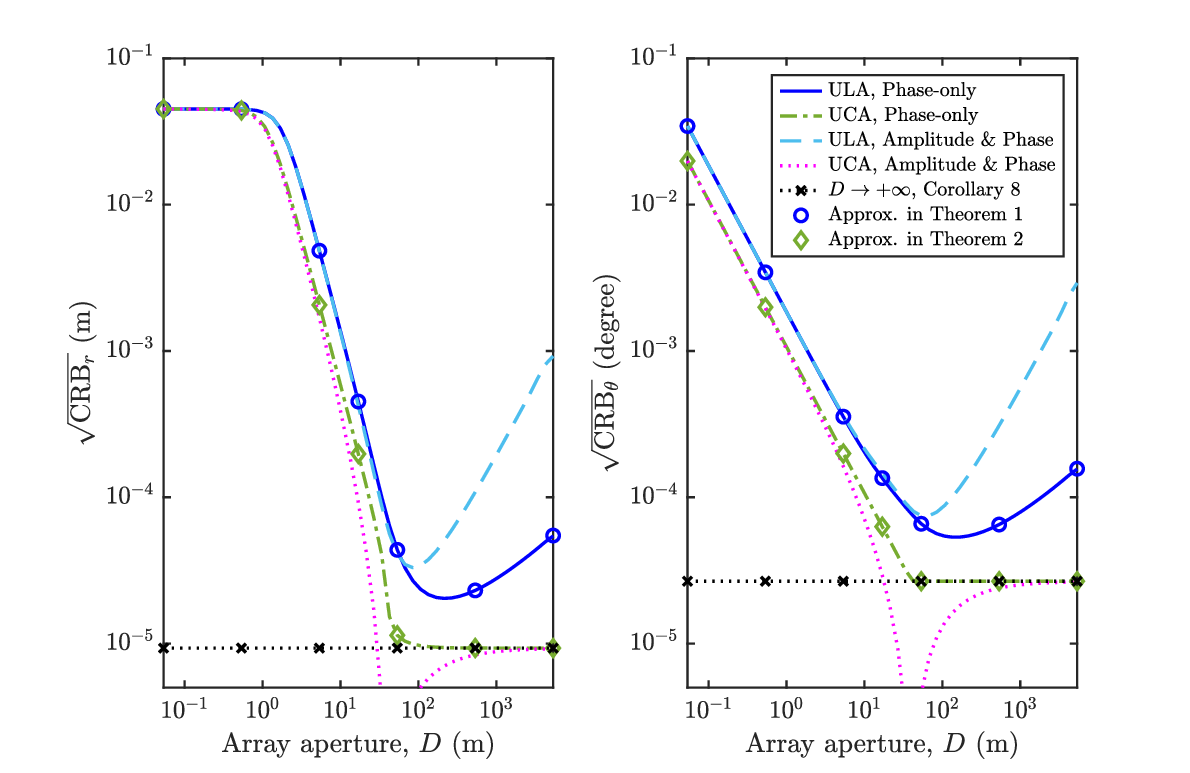}
		\caption{$\theta = \pi/4$}
	\end{subfigure}
	\begin{subfigure}{0.48\textwidth}
		\centering
		\includegraphics[width=1\textwidth]{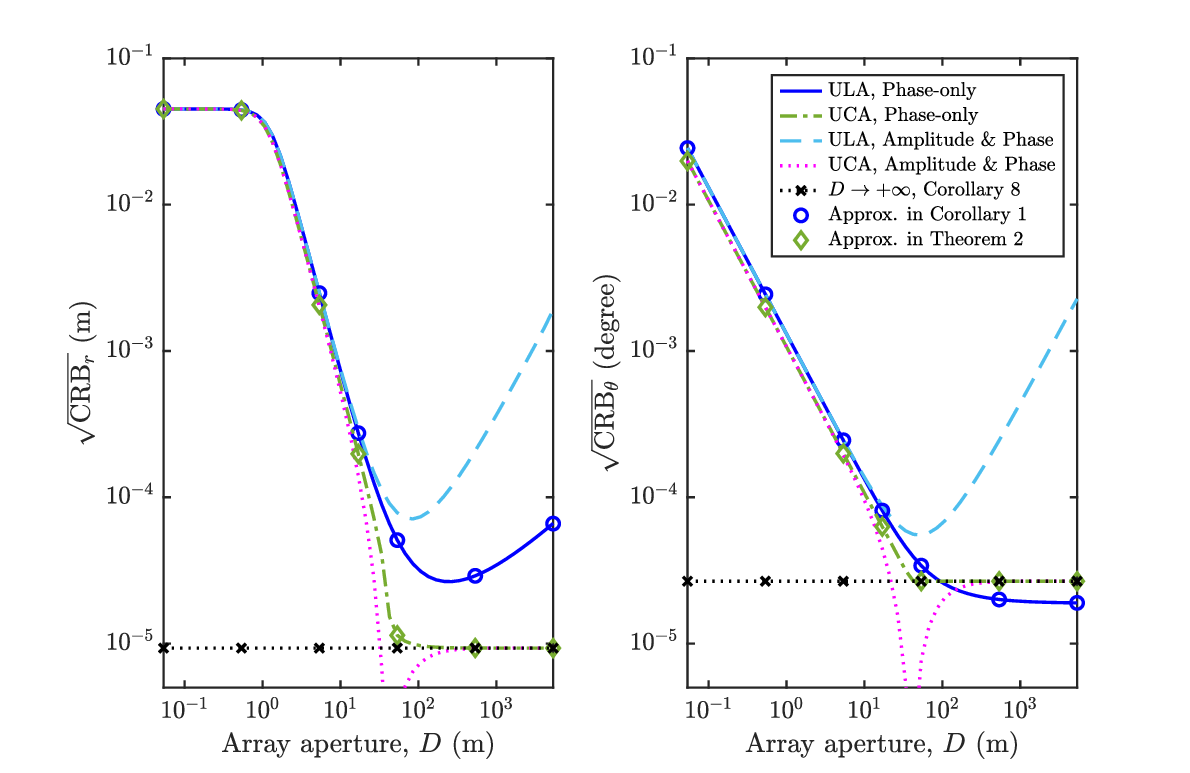}
		\caption{$\theta = \pi/2$.}
	\end{subfigure}
 \caption{CRBs versus the array aperture under the condition of a fixed number of antennas.}
 \label{fig_aperture}
\end{figure}

Fig. \ref{fig_antenna_number} examines the impact of the number of antennas $N$ on CRBs under the conditions of either fixed array aperture $D$ or fixed antenna spacing $d$. For fixed antenna spacing, it is set to $d = \lambda_c/2$ for ULAs and $d = \pi \lambda_c/2$ for UCAs to ensure the same aperture, where $\lambda_c$ is the signal wavelength corresponding to the central frequency $f_c$. It can be observed that by considering \emph{only the phase variation}, when the array aperture is fixed, the CRBs decrease linearly (on a logarithmic scale) as the number of antennas increases. This is because additional antennas provide only more observation samples in the spatial domain, as discussed in \textbf{Remark \ref{remark_3}}. Conversely, when the antenna spacing is fixed, increasing the number of antennas also enlarges the array aperture. Under these circumstances, the CRBs exhibit non-linear behavior due to the changes in the near-field and spatial-wideband effects. In particular, for the non-broadside target with $\theta = \pi/4$ as depicted in Fig. \ref{fig_antenna_number}(a), both $\mathrm{CRB}_r$ and $\mathrm{CRB}_{\theta}$ achieved by ULAs initially decrease rapidly and then converge to a limit value, as specified in \textbf{Corollary \ref{Theorem_ULA_N_limit}}, as the number of antennas increases. However, for the broadside target with $\theta = \pi/2$, $\mathrm{CRB}_{\theta}$ achieved by ULAs decreases without bounds. This result is also consistent with \textbf{Corollary \ref{Theorem_ULA_N_limit}}, since $\cos \theta = 0$ in this case. Regarding UCAs, both $\mathrm{CRB}_r$ and $\mathrm{CRB}_{\theta}$ decrease unboundedly as the number of antenna increases, regardless of the angle of the target, which is consistent with \textbf{Corollary \ref{corollary_UCA_2}}. 

\begin{figure}[t!]
	\centering
    \includegraphics[width=0.48\textwidth]{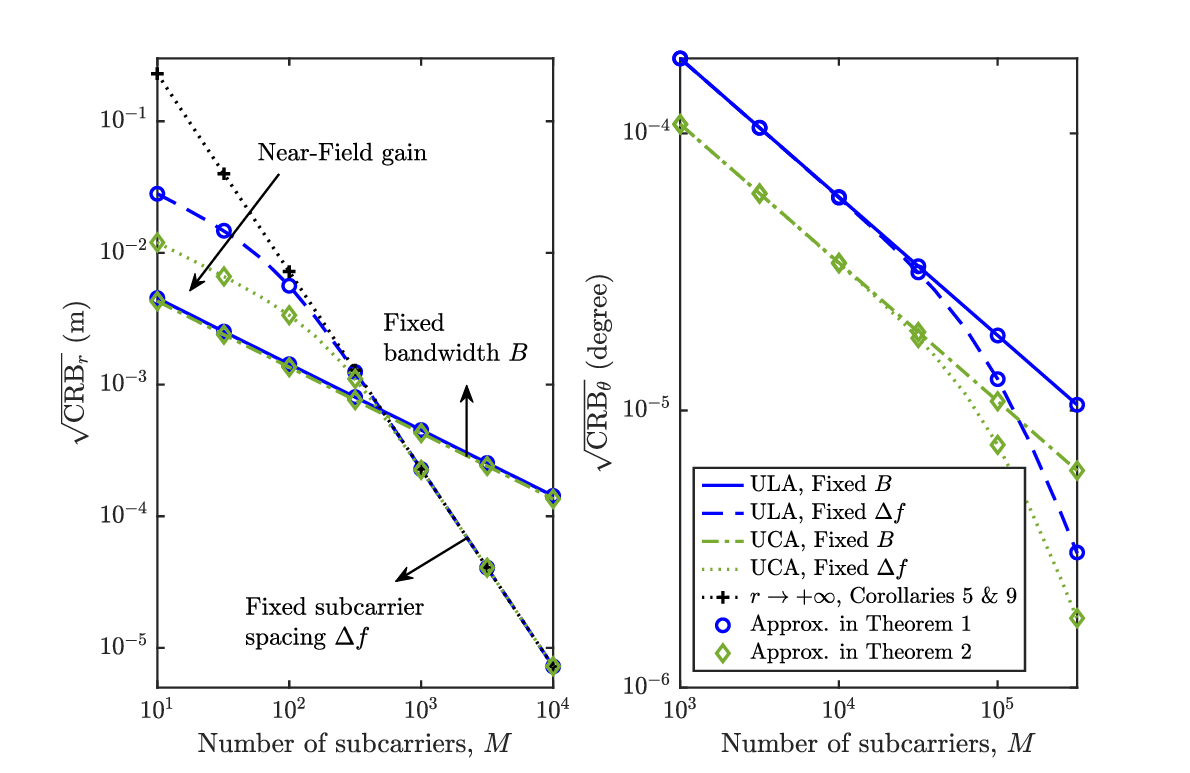}
    \caption{\textcolor{black}{CRBs versus the number of subcarriers under the conditions of fixed bandwidth and fixed subcarrier spacing, respectively.}}
	\label{fig_carrier_number}
	\vspace{-0.25cm}
\end{figure}

\begin{figure}[t!]
	\centering
    \includegraphics[width=0.48\textwidth]{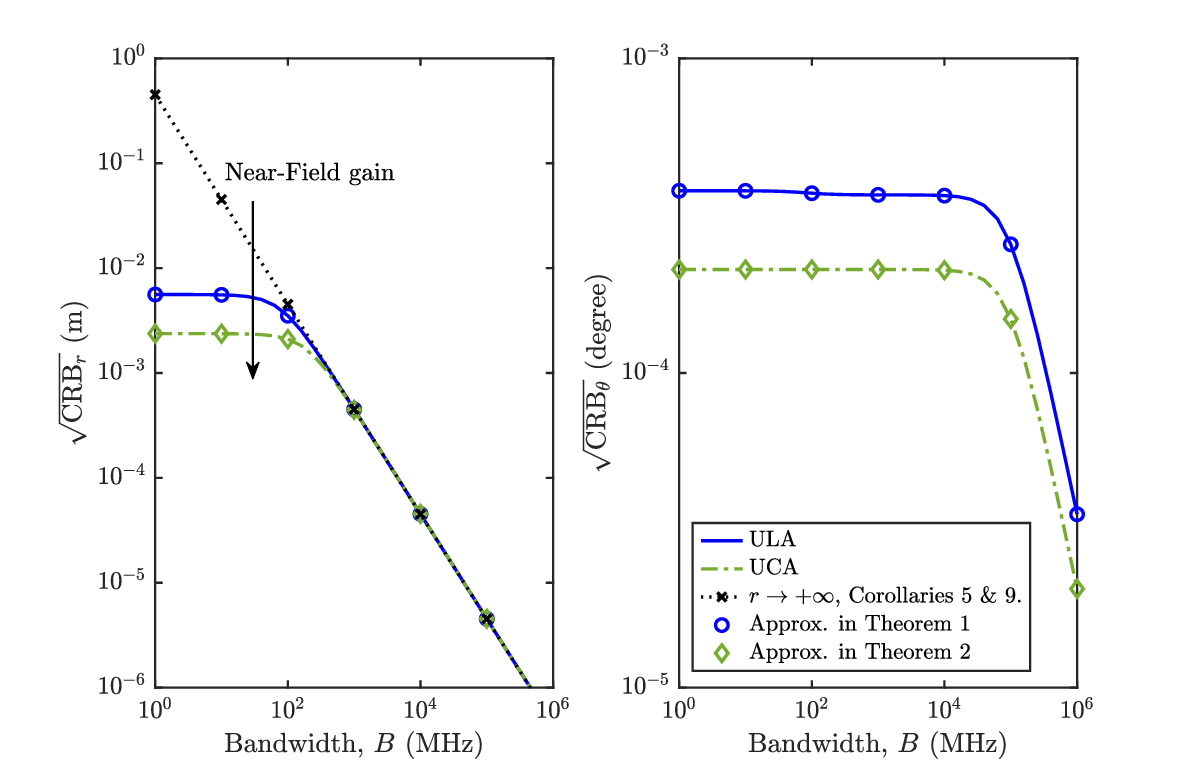}
    \caption{CRBs versus the bandwidth under the condition of a fixed number of subcarriers.}
	\label{fig_bandwidth}
\end{figure}

\textcolor{black}{We also evaluate the CRBs considering \emph{both amplitude and phase variations} in Fig. \ref{fig_antenna_number}, which are calculated using the accurate array response vector in \eqref{array_response}, the spatially white transmit covariance in \eqref{eqn_transmit_cov}, and the CRB formulas in \eqref{CRB_a} and \eqref{CRB_b}. It is interesting to observe that for ULAs, amplitude variation leads to a significant performance loss when the array aperture is enlarged by increasing the number of antennas. This is expected and consistent with our discussion in \textbf{Remark \ref{remark_6}} because the spatially white transmit covariance in \eqref{eqn_transmit_cov} becomes unsuitable under these conditions. Specifically, as the array aperture of a ULA grows, the links between the target and the antennas at the ends of the array experience a severe pathloss, contributing minimally to sensing. However, the spatially white transmit covariance continues to allocate the same power to all antennas. This approach reduces the effective power available for sensing, resulting in performance degradation. For UCAs, however, amplitude variation has a negligible impact on sensing performance. This is because significant amplitude variation only occurs when the radius of the UCA approaches the target distance.}

Fig. \ref{fig_aperture} further investigates the impact of array aperture $D$ while keeping the space-domain samples, i.e., the number of antennas $N$, constant. When considering only the phase variations, it is interesting to observe that a larger aperture does not always lead to better sensing performance, as discussed in \textbf{Remark \ref{remark_2}}. Specifically, for ULAs and a non-broadside target ($\theta = \pi/4$), an aperture larger than $10^2$ meters results in increased $\mathrm{CRB}_r$ and $\mathrm{CRB}_{\theta}$, indicating degraded sensing performance, as shown in Fig. \ref{fig_aperture}(a). This result is consistent with \textbf{Corollary \ref{theorem_ULA_D_limit}}. Furthermore, for the broadside target ($\theta = \pi/2$) depicted in Fig. \ref{fig_aperture}(b), $\mathrm{CRB}_{\theta}$ eventually converges to the limit given in \textbf{Corollary \ref{theorem_ULA_D_limit}}. For UCAs, the CRBs converge to the limit described in \textbf{Corollary \ref{corollary_UCA_3}} for both broadside and non-broadside targets. \textcolor{black}{Similarly, amplitude variation results in performance degradation.}

\begin{figure}[t!]
	\centering
    \includegraphics[width=0.42\textwidth]{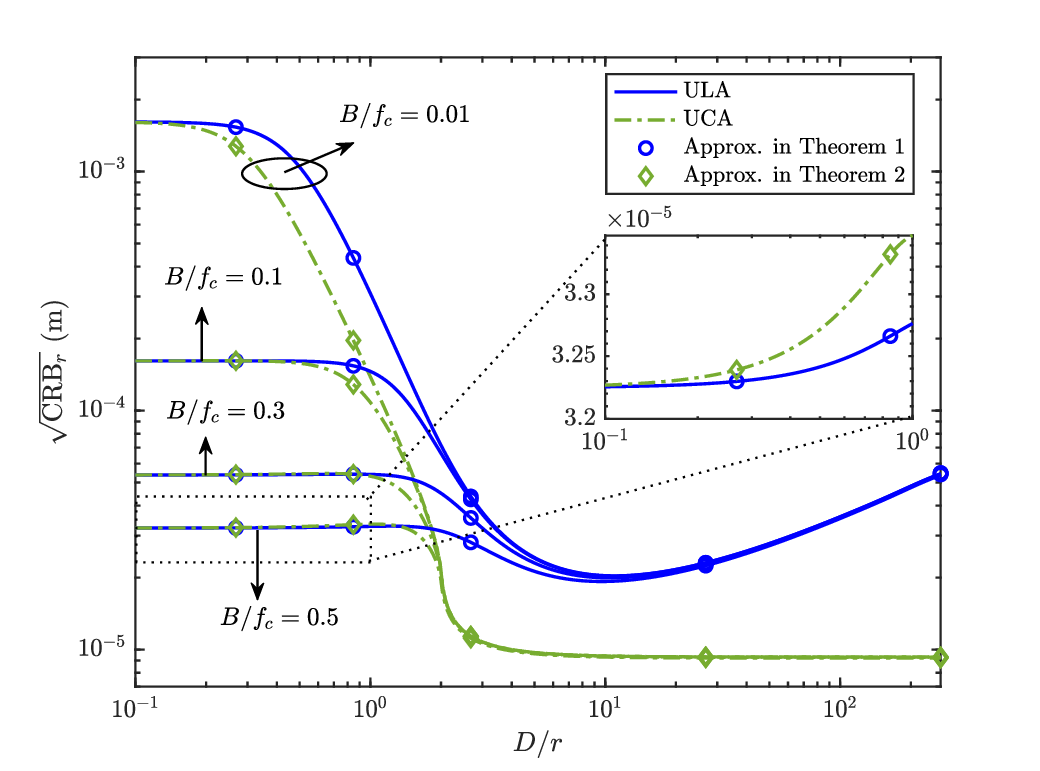}
    \caption{\textcolor{black}{CRBs versus the aperture-to-distance ratios with various bandwidth-to-carrier-frequency ratios.}}
	\label{fig_tradeoff}
	\vspace{-0.15cm}
\end{figure}

\begin{figure}[t!]
	\centering
    \includegraphics[width=0.48\textwidth]{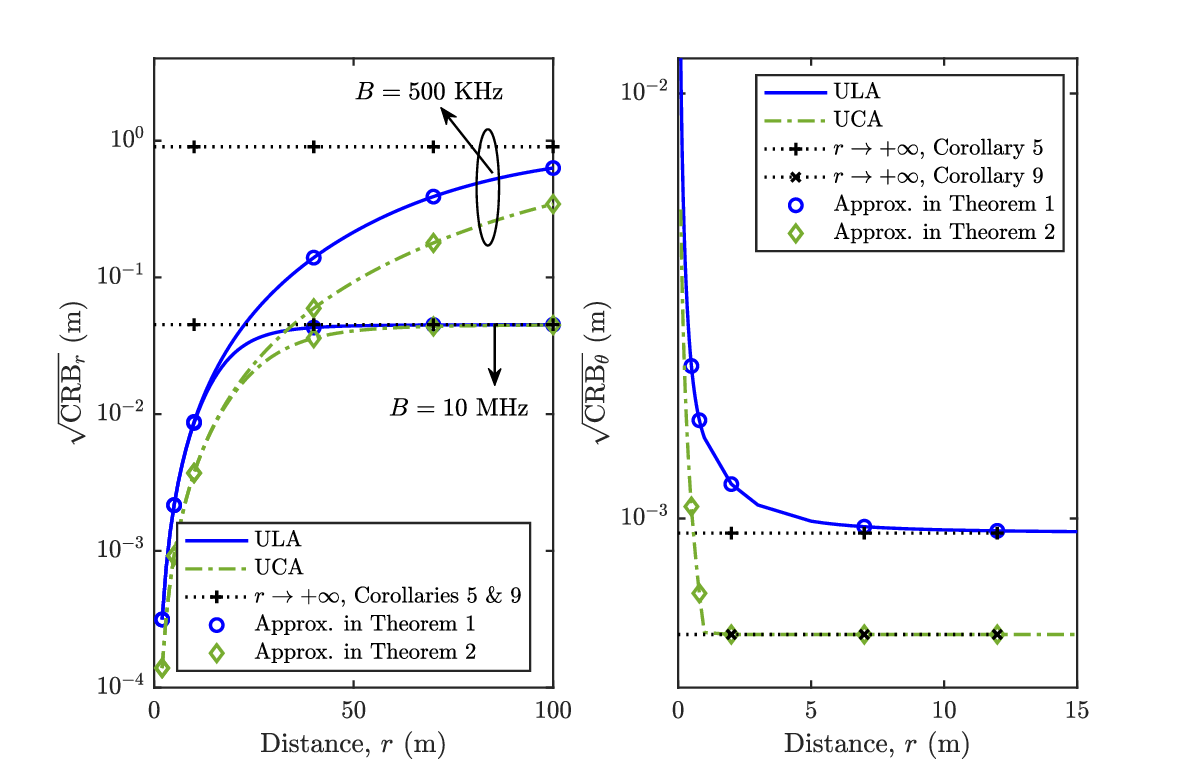}
    \caption{\textcolor{black}{CRBs versus the target distance.}}
	\label{fig_distance}
\end{figure}

\subsection{Impact of Bandwidth}
Fig. \ref{fig_carrier_number} studies the impact of the number of subcarriers $M$ on CRBs under the conditions of either fixed bandwidth $B$ or fixed subcarrier spacing $\Delta f$, respectively. Similar to the impact of the number of antennas, when the bandwidth is fixed, increasing the number of subcarriers merely increases the number of observation samples in the frequency domain, thus leading to a linear decrease of CRBs on a logarithmic scale. In the fixed subcarrier spacing scenario, increasing the number of subcarriers also expands the bandwidth. From Fig. \ref{fig_carrier_number}, it can be observed that increasing the bandwidth has a marginal influence on $\mathrm{CRB}_{\theta}$ except when the bandwidth is extremely large, which is consistent with \textbf{Remark \ref{remark_1}}. This conclusion is further confirmed in Fig. \ref{fig_bandwidth} demonstrating the effect of only bandwidth by keeping the number of subcarriers $M$ unchanged. In particular, a bandwidth larger than $10^5$ MHz, which is almost impossible in practice, is required to achieve a significant improvement in angle estimation. Furthermore, combining the results in Fig. \ref{fig_carrier_number} and Fig. \ref{fig_bandwidth}, we can conclude that the near-field gain in distance estimation is substantial when the bandwidth is small, but becomes negligible when the bandwidth is large.

\subsection{Tradeoff Between Array Size and Bandwidth}

Fig. \ref{fig_tradeoff} illustrates the joint impact of array size and bandwidth on sensing performance based on the aperture-to-distance ratio $D/r$ and the bandwidth-to-carrier-frequency ratio $B/f_c$. Since the impact of bandwidth on angle sensing is marginal, we focus solely on the performance of distance sensing. It can be observed that when $B/f_c$ is small, i.e., less than 0.1, increasing $D/r$ consistently results in better distance sensing performance when $D/r \le 10$ for both ULAs and UCAs. However, when $B/f_c$ is extremely large, e.g., $B/f_c = 0.5$, increasing $D/r$ leads to degraded sensing performance in most cases, as discussed in \textbf{Remark \ref{remark_2}}. Furthermore, when $D/r \le 1$ and $B/f_c \ge 0.1$, the performance improvement or degradation caused by increasing $D/r$ is actually negligible. This result suggests that in a practical sensing system with ultra-large bandwidth, the array aperture design should focus more on the performance of angle estimation and increase the number of space-domain observations, as distance estimation relies almost solely on the bandwidth.

\begin{figure}[t!]
	\centering
	\begin{subfigure}{0.24\textwidth}
		\centering
		\includegraphics[width=1\textwidth]{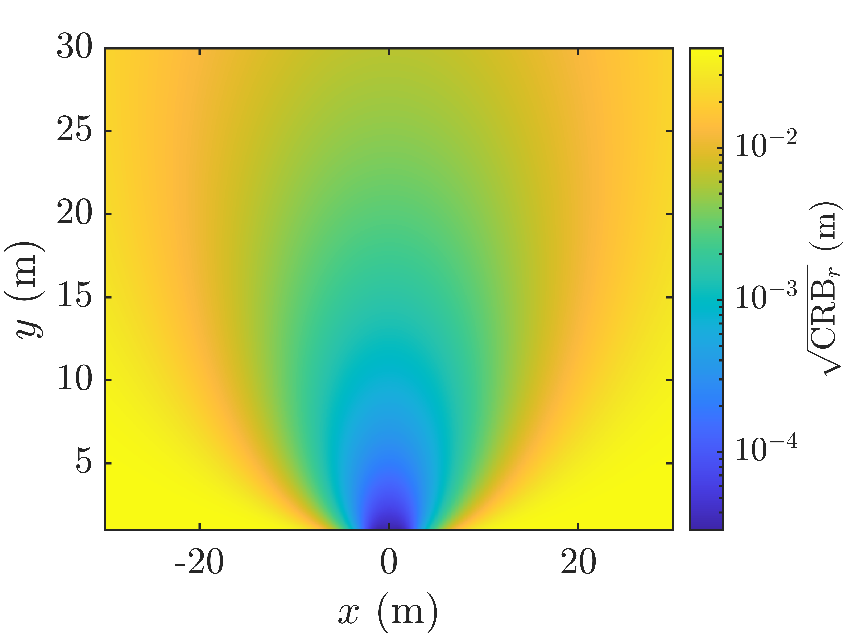}
		\caption{ULA, distance.}
	\end{subfigure}
	\begin{subfigure}{0.24\textwidth}
		\centering
		\includegraphics[width=1\textwidth]{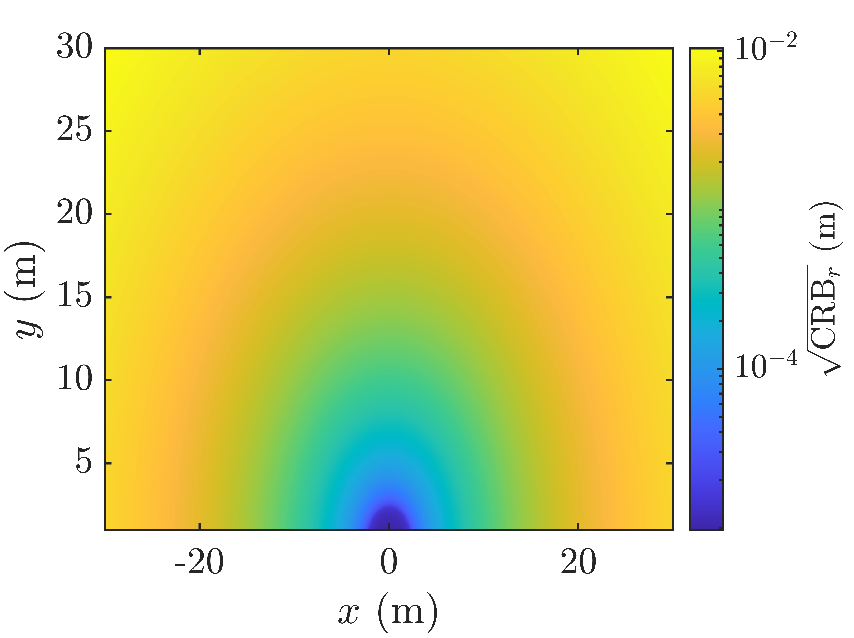}
		\caption{UCA, distance.}
	\end{subfigure}
	\begin{subfigure}{0.24\textwidth}
		\centering
		\includegraphics[width=1\textwidth]{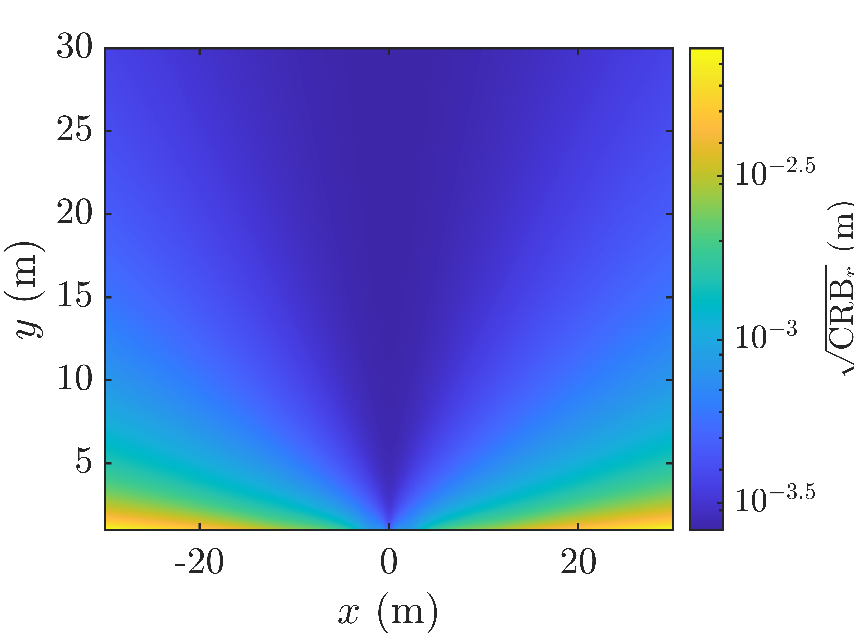}
		\caption{ULA, angle.}
	\end{subfigure}
	\begin{subfigure}{0.24\textwidth}
		\centering
		\includegraphics[width=1\textwidth]{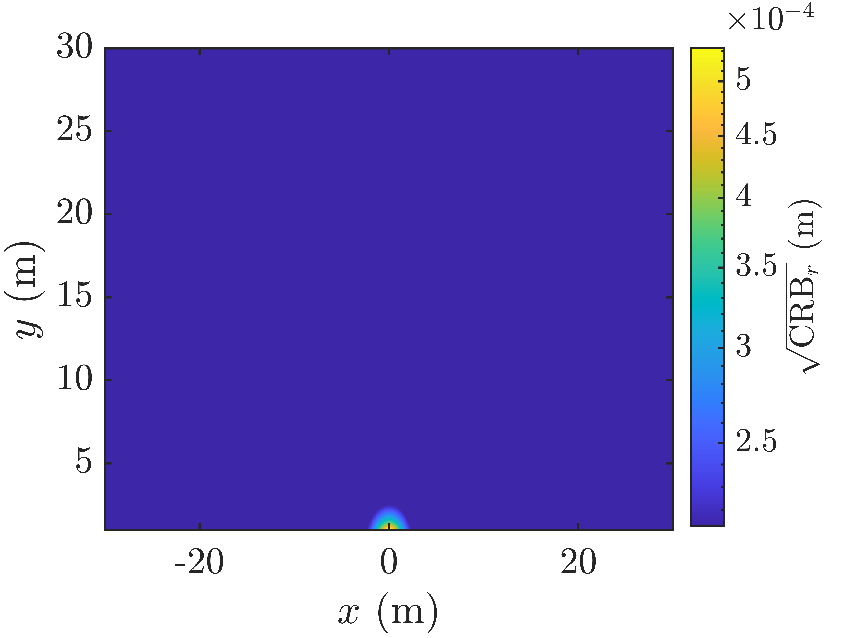}
		\caption{UCA, angle.}
	\end{subfigure}
 \caption{\textcolor{black}{CRBs achieved by ULAs and UCAs versus the location of the target.}}
 \label{fig_angle}
\end{figure}

\subsection{Impact of Target Location}

\textcolor{black}{Fig. \ref{fig_distance} explores the impact of target distance on sensing performance when $D = 2$ m. There are two key observations. First, the accuracy of distance estimation approaches the far-field bound more rapidly with larger bandwidths, indicating that a larger bandwidth reduces the extent of the near-field effect. This also highlights that the near-field effect plays a more significant role in narrowband systems compared to wideband systems. Second, regarding angle estimation, the value of $\mathrm{CRB}_{\theta}$ achieved by both ULAs and UCAs initially decreases and then converges to a stable lower bound as $r$ increases, aligning with our analysis in \textbf{Remark \ref{remark_2}}.}

Fig. \ref{fig_angle} further explores the achievable CRBs for targets at different locations. It is evident that the performance of ULAs is significantly influenced by the angle of the target. In particular, the CRBs for both distance and angle estimation achieved by ULAs increase as the target deviates from the broadside direction, indicating a degraded sensing performance. However, the sensing performance of UCAs is independent of the target's angle, as discussed in \textbf{Remark \ref{remark_7}}.

\vspace{-0.2cm}
\section{Conclusion} \label{sec:conclusion}
This study has examined the joint effects of wideband and near-field phenomena on sensing performance in an OFDM communication system, with an in-depth analysis of CRBs for angle and distance estimation. The impact of key system parameters, including array size, bandwidth, and target location, on sensing performance has been revealed, providing valuable insights for future research and applications. Additionally, our results indicate that a new transmit signal design is required for NISE when signal amplitude variation is significant, presenting an interesting direction for future research.

\begin{appendices}

\section{Proof of Theorem \ref{theorem_ULA_1}} \label{proof_theorem_ULA_1}

According to \eqref{ULA_distance}, the derivatives of $r_n$ with respect to $\theta$ and $r$ for ULAs are   
\begin{align}
    \frac{\partial r_n}{\partial \theta} = &\frac{r \chi_n d \sin \theta }{\sqrt{ r^2 + \chi_n^2 d^2 - 2 r \chi_n d \cos \theta }}, \nonumber\\ \frac{\partial r_n}{\partial r} = &\frac{r - \chi_n d \cos \theta }{\sqrt{ r^2 + \chi_n^2 d^2 - 2 r \chi_n d \cos \theta }}. 
\end{align}
Then, by defining $\delta = \frac{1}{N}$, the intermediate parameter $u_{\theta}$ can be expressed as 
\begin{align}
    u_{\theta} & = \sum_{n = -\frac{N-1}{2}}^{\frac{N-1}{2}} \frac{r^2 n^2 d^2 \sin^2 \theta }{ r^2 + n^2 d^2 - 2 r n d \cos \theta } \nonumber \\ 
    &= \sum_{n = -\frac{N-1}{2}}^{\frac{N-1}{2}} \frac{r^2 N^3 d^2 n^2 \delta^2 \sin^2 \theta}{r^2 + N^2 d^2 n^2 \delta^2 - 2 r N d n \delta \cos \theta} \delta \nonumber \\
    & \overset{(a)}{=} \int_{-\frac{1}{2}}^{\frac{1}{2}} \frac{r^2 N^3 d^2 x^2 \sin^2 \theta}{r^2 + N^2 d^2 x^2 - 2 r N d x \cos \theta} dx 
    \overset{(b)}{=} \eqref{u_t_value},
\end{align}   
where $(a)$ stems from the fact that $\delta \ll 1$ and step $(b)$ is obtained based on the integral formulas \cite[Eq. (2.175.4)]{gradshteyn2014table} and \cite[Eq. (2.172)]{gradshteyn2014table}. Additionally, the expressions of $G_1$, $G_2$, and $\Xi$ are given in \eqref{ULA_au_value}.

The closed-form expression of $u_r$ is given by 
\begin{equation}
    u_r = \sum_{n = -\frac{N-1}{2}}^{\frac{N-1}{2}} \frac{ ( r - nd\cos \theta )^2 }{ r^2 + n^2 d^2 - 2 r n d \cos \theta } = \eqref{u_r_value}.
\end{equation}
Furthermore, the closed-form expression of $c_{\theta}$ can be derived as follows:
\begin{align}
    c_{\theta} = & \int_{- \frac{1}{2}}^{\frac{1}{2}} \frac{r N^2 d x \sin \theta}{\sqrt{r^2 + N^2 d^2 x^2 - 2 N d x \cos \theta}} dx
    \overset{(c)}{=} \eqref{c_t_value},
\end{align}
where step $(c)$ is obtained by using the integral formulas \cite[Eq. (2.264.2)]{gradshteyn2014table} and \cite[Eq. (2.261)]{gradshteyn2014table}. Similarly, the expression of $c_r$ can be derived as follows:
\begin{align}
    c_r &= \int_{- \frac{1}{2}}^{\frac{1}{2}} \frac{r N}{\sqrt{r^2 + N^2 d^2 x^2 - 2 r N d x \cos \theta}} d x - \frac{\cos \theta}{r \sin \theta} c_{\theta} \nonumber\\
    &= \eqref{c_r_value}.
\end{align}
Regarding the parameter $\varepsilon$, it can be derived as follows:
\begin{align}
    \varepsilon &= \int_{- \frac{1}{2}}^{\frac{1}{2}} \frac{r^2 N^2 d x \sin \theta}{r^2 + N^2 d^2 x^2 - 2 r N d x \cos \theta} dx - \frac{\cos \theta}{r \sin \theta} u_{\theta}\nonumber \\ &\overset{(d)}{=} \eqref{eta_value},
\end{align} 
where step $(d)$ is obtained using the integral formulas \cite[Eq. (2.175.1)]{gradshteyn2014table} and \cite[Eq. (2.172)]{gradshteyn2014table}. The proof is thus completed.

\section{Proof of Corollary \ref{Theorem_ULA_N_limit}} \label{proof_Theorem_ULA_N_limit}
For fixed antenna spacing $d = D/N$, the parameter $D$ in \eqref{u_t_value}-\eqref{eta_value} needs to be replaced by $Nd$. In this case, as $N \rightarrow +\infty$, we must have $Nd \gg r$, leading to the following approximations:
\begin{align}
    &\Xi \approx \frac{\pi}{2} + \frac{\pi}{2} = \pi, \quad \ln \left( \frac{G_1}{G_2} \right) \approx 0, \\
    &\frac{\sqrt{ G_1 } - \sqrt{ G_2 }}{r} = \frac{ -2 N d \cos \theta }{\sqrt{ G_1 } + \sqrt{ G_2 } } \approx -2 \cos \theta, \\
    &\ln \left(\frac{ \sqrt{G_1} + \frac{1}{2} N d- r \cos \theta}{ \sqrt{G_2} - \frac{1}{2}N d - r \cos \theta}\right) \approx  \ln \left(\frac{ \sqrt{ 1 + \left(\frac{2r\sin \theta}{N d}\right)^2 } + 1 }{\sqrt{ 1 + \left(\frac{2r\sin \theta}{N d}\right)^2 } - 1} \right) \nonumber \\
    &\overset{(a)}{\approx}  \ln \left(\frac{1 + \left(\frac{r \sin \theta}{ N d} \right)^2}{ \left( \frac{r \sin \theta}{N d} \right)^2} \right) \approx \ln \left(\frac{N d}{ r \sin \theta} \right)^2,
\end{align} 
where step $(a)$ is derived according to the first-order Taylor series $\sqrt{1 + x^2} \approx 1 + \frac{1}{2} x^2$ for $x \approx 0$.  
Based on the above results, the intermediate parameters can be simplified as
\begin{align}
    u_{\theta} &\approx r^2 \sin^2 \theta \left( N +  \frac{\pi r \cos 2 \theta}{d \sin \theta} \right), \\
    u_{r} &\approx N \cos^2 \theta - \frac{\pi r \sin \theta \cos 2 \theta}{d}, \\
    c_{\theta} & \approx \frac{r^2 \sin 2\theta}{d}\ln\left( \frac{N d}{r \sin \theta}\right), \\ 
    c_r & \approx \frac{2 r \sin^2 \theta}{d} \ln\left( \frac{N d}{r \sin \theta}\right) , \\
    \varepsilon & \approx \frac{\pi r^2 \cos \theta}{d} - r \sin \theta \cos \theta \left(N +  \frac{\pi r \cos 2 \theta}{d \sin \theta}\right).
\end{align}
To obtain the closed-form approximation of $\mathrm{CRB}_{\theta}$ and $\mathrm{CRB}_r$ in \eqref{CRB_theta} and \eqref{CRB_r} when $N d \gg r$, we first derive the following intermediate parameters:
\begin{align}
    \phi &\triangleq u_{\theta} u_r - \varepsilon^2 \approx \frac{\pi r^3 \sin \theta}{d} \left( N - \frac{\pi r \sin\theta}{d} \right), \\
    \psi &\triangleq u_{\theta} c_r^2 + u_r c_{\theta}^2 - 2 \varepsilon c_{\theta} c_r \nonumber \\
    &\approx \frac{4 r^4 \sin^2\theta}{d^2} \left( N - \frac{\pi r \sin\theta}{d} \right) \ln^2\left( \frac{N d}{r \sin \theta}\right).
\end{align}
Then, defining $U = M M_2 - 2M_1^2$, the $\mathrm{CRB}_{\theta}$ can be approximated as follows:
\begin{align}
    \label{appendix_B_CRB_theta_approx}
    &\mathrm{CRB}_{\theta} = \frac{ N M M_2 u_{r} + U c_{r}^2 }{ 4 \rho L \left( N M M_2^2 \phi + U M_2 \psi \right)} \nonumber \\
    &\approx \frac{M M_2 \left(\cos^2\theta - \frac{\pi r \sin \theta \cos 2\theta}{Nd} \right) + \frac{4r^2 \sin^4 \theta}{N^2 d^2} U \ln^2 \left( \frac{Nd}{r\sin\theta} \right) }{ 4 \rho L \left( \splitdfrac{\frac{\pi r^3 \sin \theta}{d} M M_2^2 \left( 1 - \frac{\pi r \sin \theta}{Nd} \right) }{+ \frac{4 r^4 \sin^2\theta}{N^2 d^2} U M_2 \left( N - \frac{\pi r \sin \theta}{d} \ln^2 \left( \frac{Nd}{r \sin \theta} \right) \right)} \right) } \nonumber \\
    &\overset{(b)}{\approx} \frac{d \cos^2 \theta}{4 \rho L M_2 \pi r^3 \sin \theta },
\end{align} 
where step $(b)$ is obtained based on the limits $\frac{1}{x} \rightarrow 0$, $\frac{1}{x} \ln^2x \rightarrow 0$, and $\frac{1}{x^2} \ln^2x \rightarrow 0$ when $x \rightarrow +\infty$. Following a similar process, we have 
\begin{equation} \label{appendix_B_CRB_r_approx}
    \mathrm{CRB}_{r} \approx \frac{d \sin \theta}{4 \rho L M_2 \pi r}.
\end{equation}
Based on the above approximations, the limits in \textbf{Corollary \ref{Theorem_ULA_N_limit}} can be readily obtained. 
The proof is thus completed.

\section{Proof of Corollary \ref{theorem_ULA_D_limit}} \label{proof_theorem_ULA_D_limit}

We first derive the asymptotic value of $\mathrm{CRB}_{\theta}$. When $\theta = \frac{\pi}{2}$, $\mathrm{CRB}_{\theta}$ can be approximated by \eqref{CRB_theta_ULA_approx}, where the impact of $D$ is reflected solely in the function $\Phi \left( \frac{D}{r} \right)$. It is easy to show that $\lim_{D \rightarrow +\infty} \Phi \left( \frac{D}{r} \right) = 1$.
Thus, according to \eqref{CRB_theta_ULA_approx}, we have 
\begin{equation}
    \lim_{D \rightarrow +\infty} \mathrm{CRB}_{\theta} = \frac{3}{\rho L N M r^2 (12 f_c^2 + B^2 - \Delta f^2) }.
\end{equation}
When $\theta \neq \frac{\pi}{2}$, according to \eqref{appendix_B_CRB_theta_approx}, we have
\begin{align}
    \lim_{D \rightarrow +\infty} \mathrm{CRB}_{\theta} = &\lim_{D \rightarrow +\infty} \frac{d \cos^2 \theta}{4 \rho L M_2 \pi r^3 \sin \theta} \nonumber \\ 
    = &\lim_{D \rightarrow +\infty} \frac{D \cos^2 \theta}{4 \rho L N M_2 \pi r^3 \sin \theta} = +\infty.
\end{align}
Similarly, the limit of $\mathrm{CRB}_r$ can be calculated based on \eqref{appendix_B_CRB_r_approx} as follows:
\begin{align}
    \lim_{D \rightarrow +\infty} \mathrm{CRB}_r = \lim_{D \rightarrow +\infty} \frac{D \sin \theta}{4 \rho L N M_2 \pi r} = +\infty.
\end{align}
The proof is thus completed.

\section{Proof of Corollary \ref{theorem_ULA_far}} \label{proof_theorem_ULA_far}

\textcolor{black}{When $r \rightarrow +\infty$, the far-field approximation in \eqref{far-field_distance} is achieved with strict equality as $r_n = r - n d \cos \theta$.    
Then, its partial derivatives with respect to $r$ and $\theta$ become $\frac{\partial r_n}{\partial \theta} = n d \sin\theta$ and $\frac{\partial r_n}{\partial r} = 1$, respectively. 
Following Appendix \ref{proof_theorem_ULA_1}, when $r \rightarrow +\infty$ and $N \gg 1$, we have
\begin{align}
    \label{appx_c_1}
    u_{\theta} &= \sum_{n = -\frac{N-1}{2}}^{\frac{N-1}{2}} n^2d^2\sin^2 \theta = \frac{N D^2 \sin^2 \theta}{12}, \\
    u_r &= \sum_{n = -\frac{N-1}{2}}^{\frac{N-1}{2}} 1 = N, \quad
    c_{\theta} = \sum_{n = -\frac{N-1}{2}}^{\frac{N-1}{2}} n d \sin\theta = 0, \\
    c_r &= \sum_{n = -\frac{N-1}{2}}^{\frac{N-1}{2}} 1 = N, \quad
    \varepsilon = \sum_{n = -\frac{N-1}{2}}^{\frac{N-1}{2}} n d \sin\theta = 0.
\end{align}
The results in \textbf{Corollary \ref{theorem_ULA_far}} can be obtained by substituting the above results into \eqref{CRB_theta} and \eqref{CRB_r}. The proof is thus completed.}

\section{Proof of Theorem \ref{theorem_UCA}} \label{proof_theorem_UCA}

For UCAs, partial derivatives of the propagation distance $r_n$ with respect to $r$ and $\theta$ are given by
\begin{align}
    \frac{\partial r_n}{\partial \theta} = \frac{r R \sin(\theta - \frac{2\pi n}{N})}{\sqrt{ r^2 + R^2 - 2 r R \cos( \theta - \frac{2\pi n}{N}) }}, \\
    \frac{\partial r_n}{\partial r} = \frac{r - R \cos(\theta - \frac{2\pi n}{N})}{\sqrt{ r^2 + R^2 - 2 r R \cos( \theta - \frac{2\pi n}{N}) }}.
\end{align}
By defining $\delta = \frac{2\pi}{N}$, the parameter $u_{\theta}$ can be derived as     
\begin{align}
    \label{AD_1}
    u_\theta & = \sum_{n=1}^N \left( \frac{\partial r_n}{\partial \theta} \right)^2 = \frac{r^2 R^2}{\delta} \sum_{n=1}^{N} \frac{\sin^2(\theta - n \delta)}{r^2 + R^2 - 2 r R \cos(\theta - n \delta)} \delta \nonumber \\
    &\overset{(a)}{\approx} \frac{r^2 R^2 N}{2\pi} \int_{0}^{2 \pi} \frac{\sin^2 x}{r^2 + R^2 - 2 r R \cos x} dx \nonumber \\
    & = \frac{r^2 R^2 N}{2 \pi} \Bigg( \int_{0}^{\pi} \frac{\sin^2 x}{r^2 + R^2 - 2 r R \cos x} dx \nonumber \\ & + \int_{0}^{\pi} \frac{\sin^2 x}{r^2 + R^2 + 2 r R \cos x} dx \Bigg) \overset{(b)}{=} \begin{cases}
        \frac{r^2 N}{2}, & R \ge r \\
        \frac{R^2 N}{2}, & R < r
    \end{cases},
\end{align}
where approximation $(a)$ is obtained based on $\delta \ll 1$ when $N \gg 1$ and step $(b)$ is derived based on the integral formula \cite[Eq. (3.613.3)]{gradshteyn2014table}. Similarly, the remaining parameters can be derived as follows:    
\begin{align}
    u_r &= N - \frac{1}{r^2} u_{\theta}, \\
    c_{\theta} & \approx \int_{0}^{2 \pi} \frac{r R N \sin x}{2 \pi \sqrt{ r^2 + R^2 - 2r R_s \cos x }} dx \overset{(c)}{=} 0, \\
    c_r & \approx \int_{0}^{2\pi} \frac{N (r - R \cos x)}{ 2\pi \sqrt{r^2 + R^2 - 2 r R \cos x} } dx = N \Upsilon\left(\frac{r}{R}\right), \\
    \label{AD_2}
    \varepsilon &\approx \int_{0}^{2 \pi} \frac{ r R N (R \sin x \cos x -r \sin x)}{2\pi(r^2 + R^2 - 2 r R \cos x)} dx \overset{(d)}{=} 0,
\end{align}
where steps $(c)$ and $(d)$ are obtained according to the symmetry property of the functions and function $\Upsilon(\alpha)$ is given by 
\vspace{-0.15cm}
\begin{equation}
    \Upsilon(\alpha) = \int_{0}^{2\pi} \frac{\alpha - \cos x}{ 2\pi \sqrt{1 - 2\alpha \cos x + \alpha^2}} dx.
\end{equation}
It can be proved that the function $\Upsilon(\alpha)$ is a transcendental function that does not have a closed-form expression. Substituting \eqref{AD_1}-\eqref{AD_2} into \eqref{CRB_theta} and \eqref{CRB_r}, the results in \textbf{Theorem \ref{theorem_UCA}} can be obtained. The proof is thus completed.

\end{appendices}

\balance
\bibliographystyle{IEEEtran}
\bibliography{reference/mybib}


\end{document}